\documentclass[12pt,a4paper]{article}
\pdfoutput=1
\usepackage{amsmath,amssymb}
\usepackage[pdftex]{graphicx}
\usepackage{mathptmx}

\newcommand{\ignore}[1]{}

\newcommand{\keywords}[1]{{{\bf Keywords.~}{\em #1}}}

\newenvironment{program}
{
\begin{list}
{} 
{  
\setlength{\leftmargin}{20pt}
\setlength{\rightmargin}{0pt} 
\setlength{\parsep}{0pt}
\setlength{\itemsep}{0pt}
\setlength{\topsep}{0pt}
\setlength{\parskip}{0pt}
\setlength{\baselineskip}{14pt}
\setlength{\labelsep}{3pt}
}
}
{\end{list}}
\def\bp{\begin{program}}
\def\ep{\end{program}}

\renewenvironment{itemize}
{
\begin{list}
{$\bullet$}
{
\setlength{\leftmargin}{25pt}
\setlength{\rightmargin}{0pt}
\setlength{\parsep}{0pt}
\setlength{\itemsep}{0pt}
\setlength{\topsep}{\parskip}
\setlength{\parskip}{0pt}
\setlength{\baselineskip}{14pt}
\setlength{\labelsep}{3pt}
}
}
{\end{list}}

\newcounter{i}
\newenvironment{enumerat}
{
\setcounter{i}{0}
\begin{list}
{\arabic{i}.}
{
\usecounter{i}
\setlength{\leftmargin}{25pt}
\setlength{\parsep}{0pt}
\setlength{\itemsep}{0pt}
\setlength{\topsep}{\parskip}
\setlength{\parskip}{0pt}
\setlength{\baselineskip}{14pt}
\setlength{\labelsep}{5pt}
\setlength{\itemindent}{0pt}
}
}
{\end{list}}
  
\def\real{\mathbb R}
\def\diag{{\rm diag}}

\def\beq{\begin{equation}}
\def\enq{\end{equation}}
\def\bi{\begin{itemize}}
\def\ei{\end{itemize}}
\def\be{\begin{enumerat}}
\def\ee{\end{enumerat}}
\def\time{\!\times\!}

\def\real{\mathbb R}
\def\diag{{\rm diag}}

\def\beq{\begin{equation}}
\def\eeq{\end{equation}}

\def\half{\scriptstyle \frac{1}{2}}
\def\noconv{\multicolumn{2}{|c|}{no conv.}}

\begin{document}

\baselineskip 14pt

\title{\vspace{-1.5cm}
\bf\Large
Geometric scaling: a simple preconditioner for certain\\
linear systems with discontinuous coefficients}

\author{Dan Gordon\thanks{Corresponding author. 
	Dept.\ of Computer Science,
        University of Haifa, Haifa 31905, Israel.
        \newline\hspace*{18pt}{\tt gordon@cs.haifa.ac.il}} 
	\and
        Rachel Gordon\thanks{Dept.\ of Aerospace Engineering,
        The Technion--Israel Inst.\ of Technology, Haifa 32000,
        Israel.\newline\hspace*{18pt}{\tt rgordon@tx.technion.ac.il}}}

\date{April 17, 2009}

\maketitle
\vspace{-1cm}
\begin{abstract}\noindent
Linear systems with large differences between coefficients 
(``discontinuous coefficients'') arise in many cases in which 
partial differential equations (PDEs) model physical phenomena 
involving heterogeneous media.  The standard approach to solving 
such problems is to use domain decomposition (DD) techniques, 
with domain boundaries conforming to the boundaries between the 
different media.  This approach can be difficult to implement 
when the geometry of the domain boundaries is complicated or 
the grid is unstructured.  This work examines the simple 
preconditioning technique of scaling the equations by dividing 
each equation by the $L_p$-norm of its coefficients.  This 
preconditioning is called geometric scaling (GS).  
It has long been known that diagonal scaling can be useful in 
improving convergence, but there is no study on the general 
usefulness of this approach for discontinuous coefficients.  
GS was tested on several nonsymmetric linear systems with 
discontinuous coefficients derived from convection-diffusion 
elliptic PDEs with small to moderate convection terms.
It is shown that GS improved the convergence properties of 
restarted GMRES and Bi-CGSTAB, with and without the ILUT 
preconditioner.
GS was also shown to improve the distribution of the 
eigenvalues by reducing their concentration around the 
origin very significantly.
\end{abstract}
\keywords{Bi-CGSTAB; diagonal scaling; discontinuous coefficients; 
domain decomposition; geometric scaling; GMRES; GS; Lp-norm; linear 
equations; nonsymmetric equations; parallel processing; partial 
differential equations.}


\section{Introduction}

Many physical phenomena are modeled by partial differential equations 
(PDEs) which describe the relations between one or more scalar or vector 
fields and the surrounding media.  When boundary conditions are prescribed, 
a common approach to achieving a numerical solution is to impose a grid 
and discretize the equations, thus getting a system of linear equations.  
In some cases, this approach yields a system of equations with very large 
differences between coefficients of the equations.  Examples of such 
systems arise in modeling flow through heterogeneous media with 
widely-varying permeability, oil reservoir simulation, electromagnetics 
and semiconductor modeling.  Such systems are commonly referred to as 
systems with ``discontinuous coefficients''.  

One of the most common methods for tackling such problems is 
the domain-decomposition (DD) approach, in which the domain 
is partitioned into subdomains, with the subdomain boundaries 
conforming to the boundaries between the different media.  
DD techniques typically operate as follows: some boundary 
conditions are assumed to exist on the interfaces between 
subdomains, and a solution of the equations in each subdomain 
is obtained, often with only low accuracy.  The boundary 
conditions at the interfaces are then updated according to 
these solutions, and the process is repeated until convergence 
is achieved.  There exists a vast literature on this subject, 
and a review of the area is beyond the scope of this work; 
see, for example, \cite{Glowinski88,Smith96,Quarteroni99,Rice00}.  
DD techniques may be difficult to implement on unstructured 
grids or when the boundaries between domains have a 
complicated geometry.

We consider nonsymmetric linear systems with discontinuous 
coefficients derived from convection-diffusion elliptic PDEs 
with small to moderate convection terms.  It is shown that a 
simple preconditioning technique, which we call {\em geometric 
scaling} (GS), can be applied to the system of equations in 
order to improve the convergence properties of algorithms
applied to the system.  GS($p$), with an integer parameter 
$p\ge 1$, consists of dividing each equation by the $L_p$-norm 
of its vector of coefficients.  GS, which is a particular form 
of diagonal scaling on the left, is geometric in the following
sense: after applying GS($p$) to a linear system, all the rows
of the system matrix have a unity $L_p$-norm.

In order to examine the general usefulness of GS in conjunction with
various solution methods, we tested it with the two leading Krylov 
subspace solvers for nonsymmetric systems: GMRES \cite{Saad86} and 
Bi-CGSTAB \cite{Vorst92}.  Both algorithms were tested with and 
without the ILUT preconditioner \cite{Saad94} on several nonsymmetric 
problems taken from (or based on) Saad \cite{Saad03}, van der Vorst
\cite{Vorst92}, and Graham and Hagger \cite{Graham99}.  Four basic 
problems were considered, as well as several variations of the 
problems, such as modifications of the differences between the 
coefficients and various grid sizes.  Results are provided for 
three different levels of accuracy.  

Our experiments used the $L_2$-norm, but the $L_1$-norm yielded
similar results.  Our results can be characterized as follows:
\bi
\item In most cases, when the tested method (algorithm/preconditioner
combination) converges to the specified accuracy criterion, GS speeds
up the convergence time.
\item In many cases, when the tested method stagnates or diverges
on the original system, it converges on the scaled system.
\item GS was not particularly useful on problems without discontinuous
coefficients.
\item GS was not helpful on problems derived from PDEs with large
convection terms.  Such PDEs produce linear systems with very large
off-diagonal elements, and these require other solution methods.
\ei

We also provide data on the effect of GS on the distribution of the 
eigenvalues.  It is generally considered to be detrimental to the
convergence of Krylov subspace methods to have a large accumulation
of eigenvalues near the origin.
In the cases examined here, there is a large accumulation of 
eigenvalues near the origin, and GS appears to ``push'' many 
eigenvalues away from the origin.  Also, the eigenvalues of 
the scaled system appear to be distributed similarly to those 
obtained without scaling but with equal-sized coefficients.  

Note that the above choice of algorithm/preconditioner combinations 
does not imply that these methods are optimal for the above problems. 
Finding an optimal method is problem-dependent and a topic for
further research.

Diagonal scaling is not new, and its usefulness for discontinuous
coefficients has also been noted.  Van der Sluis \cite{Sluis69} 
deals with the effect of scaling on the condition number of matrices.  
Widlund \cite[p.\ 34-35]{Widlund71} writes that well-scaled ADI 
methods give good rates of convergence when the coefficients of 
elliptic problems vary very much in magnitude (ADI stands for the 
alternating direction implicit method \cite{Peaceman55}).  Graham 
and Hagger \cite[p.\ 2042-2043]{Graham99} write that diagonal 
scaling has been observed in practical computations to be very 
effective as a preconditioner for problems with discontinuous 
coefficients.  Duff and van der Vorst \cite{Duff98} write that 
on vector machines, diagonal scaling is often competitive with 
other approaches.  Regarding diagonal scaling of symmetric matrices, 
see Meurant \cite[Th. 8.1, 8.2]{Meurant99}, who also has some 
reservations about the usefulness of diagonal scaling for parallel 
processing \cite[p.\ 401]{Meurant99}.  Gambolati et al.\ 
\cite{Gambolati03} use the least square logarithm (LSL) scaling 
on the rows and the columns of the system matrix for a certain 
problem in geomechanics with discontinuous coefficients.

However, most previous works are concerned with symmetric systems
and diagonal scaling refers to multiplying the system matrix on 
the left and on the right by the same diagonal matrix, in order 
to preserve symmetry.  Also, there is no study on the general 
usefulness of geometric scaling for nonsymmetric problems with 
discontinuous coefficients.  The idea of geometric scaling was 
found to be useful for certain problems in image reconstruction 
from projections; see \cite{Gordon07}.

DD methods are often mentioned in connection with parallel 
processing.  The equations of different domains can, in 
principle, be solved in parallel by different processors.  
However, the different domains that arise in many practical 
situations do not necessarily lead to an optimal load-balancing 
assignment of equations to processors.  With the GS approach, 
inherently parallel algorithms such as Bi-CGSTAB and GMRES can 
be implemented efficiently in parallel without regard to 
subdomain boundaries.  Needless to say, GS is inherently 
parallel.

The rest of this paper is organized as follows.  \S \ref{back} 
presents some essential background.  \S \ref{prob1}--\S \ref{prob4} 
deal with the four different problems, and \S \ref{conclusions} 
concludes with a summary and some future research directions.


\section{General background}
\label{back}

Throughout the rest of the paper, we assume that all vectors are column
vectors, and we use the following notation:  $\langle p,\, q\rangle$ 
denotes the dot product of two vectors $p$ and $q$, which is also 
$p^T\!q$.  Given a vector $x=(x_1,\ldots,x_n)^T\in\real^n$, we denote 
its $L_p$-norm by $\|x\|_p = (x_1^p+\cdots+x_n^p)^{1/p}$.  
For $p=2$, we will omit the index and just write 
$\|x\|=\|x\|_2=\sqrt{\langle x,x\rangle}$.
If $A$ is an $n\time n$ matrix, we denote by $a_i$ the $i$th 
row-vector of $A$; i.e., $a_i = (a_{i1},\ldots,a_{in})^T$.  

Consider a system of $n$ linear equations in $n$ variables:
\begin{equation}
\sum_{j=1}^n a_{ij}x_j ~=~ b_i \mbox{~~for~~} 1\le i\le n
\mbox{,~~or, in matrix form:~~} Ax~=~b.
\label{linear}
\end{equation}

We shall assume throughout that (\ref{linear}) is consistent and that 
$A$ does not contain a row of zeros.  For $p\ge 1$, we define a diagonal 
matrix $D = \diag(1/\|a_1\|_p,\ldots,1/\|a_n\|_p)$.  The 
geometrically-scaled system is defined as 
\begin{equation}
 D A x ~=~ D b.
\label{scaled}
\end{equation}

In some algorithms, GS($p$) is an inherent step in the following 
sense: either the scaling is executed at the beginning as an 
intrinsic part of the algorithm, or, executing the algorithm 
produces identical results to those obtained when GS($p$) is 
done at the beginning.  As an example, it is easy to see that 
GS(2) is inherent in the Kaczmarz algorithm (KACZ) 
\cite{Kaczmarz37}.  KACZ can be described geometrically as 
follows: starting from an arbitrary point $x^0 \in \real^n$ 
as the initial iterate, KACZ projects the current iterate 
orthogonally towards a hyperplane defined by one of the 
equations.  The hyperplanes are chosen in cyclic order.  

The tests were run using the AZTEC software library \cite{Tuminaro99}, 
which includes a wide range of algorithms and preconditioners, suitable 
for sequential and parallel implementations.  Geometric scaling with 
the $L_1$-norm is a built-in option in AZTEC (where it is called 
``row-sum scaling'').  As mentioned we used Bi-CGSTAB and GMRES, 
with and without ILUT.  Restarted GMRES was used with Krylov 
subspace size of 10.  In AZTEC, GMRES is implemented with a double 
classical Gram-Schmidt orthogonalization step.  ILUT was implemented 
with AZTEC's default parameters of drop tolerance = 0 and fill-in = 1.  
These parameters are not necessarily optimal for the tested examples, 
but since our purpose was to demonstrate the general usefulness of GS, 
we stuck with a commonly-used restart value for GMRES and AZTEC's 
default values for ILUT.  No doubt, better convergence properties and 
runtimes would be obtained if these parameters were fine-tuned to each 
problem.  The eigenvalue computations were done with the LAPACK linear 
algebra package \cite{lapack}.

We also experimented with the option of multiplying the system 
matrix $A$ by $D^{\half}$ on the left and on the right, resulting 
in the system $D^{\half} A D^{\half} y = D^{\half} b$, with 
$x = D^{\half} y$.  In one problem, the results were somewhat
poorer than those obtained with GS, but otherwise, the results
were similar.  Note that this option is computationally more
expensive.
Gambolati et al.\ \cite{Gambolati03} remark that when using ILU(0),
the iteration matrix of Bi-CGSTAB on a diagonally scaled system
(on the left and/or the right) is theoretically the same as the
one with ILU(0) on the original matrix.  In our experiments, the
effect of ILUT was identical to that of ILU(0), so in theory, 
Bi-CGSTAB with ILUT on the original and the scaled systems should 
have behaved similarly.  However, our experiments indicated that 
using GS produces much better numerical results, and this also 
holds for GMRES with ILUT.  This is probably due to the fact that 
the scaled system does not have very large differences between 
coefficients, so it is much less prone to roundoff errors.


\subsection{Setup of the numerical experiments}
\label{setup}

In two dimensions, the general form of the second-order differential 
equations in this study was
\[ \frac{\partial}{\partial x}(a(x,y) u_x) +
   \frac{\partial}{\partial y}(b(x,y) u_y) + 
   \cdots ~=~ F,
\]
where $a$ and $b$ are given functions of $x$ and $y$, ``$\cdots$'' 
stands for lower-order derivatives, and $F$ is a prescribed RHS.  
In three dimensions, there are three given functions $a,b,c$ of 
$x,y,z$, and a second order partial derivative w.r.t.\ $z$ was 
also included.  Boundary conditions were either Dirichlet or mixed 
Dirichlet and Neumann.  The regions were taken as the unit square 
or the unit cube.  The discretization of the second-order 
derivatives at a given grid point $(i,j)$ was done using central 
differences; e.g., 
$\frac{\partial}{\partial x}(a u_x)$ was approximated as
\begin{eqnarray*}
\frac{\partial}{\partial x}(a u_x)_{i,j} 
& = & \left((au_x)_{i+\half,j} - (au_x)_{i-\half,j}\right) / \Delta x \\[5pt]
& = & \left(a_{i+\half,j}(u_{i+1,j}-u_{i,j}) / \Delta x -
            a_{i-\half,j}(u_{i,j}-u_{i-1,j}) / \Delta x \right) / \Delta x \\[5pt]
& = & \left( -(a_{i+\half,j} + a_{i-\half,j}) u_{i,j} +
      a_{i+\half,j} u_{i+1,j} + a_{i-\half,j} u_{i-1,j} \right) / \Delta x^2 
\end{eqnarray*}
All problems were discretized with equally-spaced grids, and the initial
value was taken as $u^0=0$.
The tests were run on a Pentium IV 2.8GHz processor with 3GB memory,
running Linux. 


\subsection{Stopping tests}

There are several stopping criteria which one may apply to iterative
systems.  Our stopping criterion was to use the relative residual:
$\|b-Ax\| / \|b-Ax^0\| < \epsilon$, where $\epsilon$ was taken as
$10^{-4}$, $10^{-7}$ and $10^{-10}$.  In some of the cases, this was
not attainable.  Since this stopping criterion depends on the scaling
of the equations, we always made this test on the geometrically-scaled
system using the $L_2$-norm.
In order to limit the time taken by the methods implemented in AZTEC, 
the maximum number of iterations was set to 10,000.  The AZTEC library 
has several other built-in stopping criteria:  numerical breakdown, 
numerical loss of precision and numerical ill-conditioning.
In the results, we denote the relative residual by rel-res, and 
non-convergence by ``no conv.''

One should note that the test for numerical breakdown in AZTEC uses 
the machine precision {\tt DBL\_EPSILON}, and this may result in a
premature notice of numerical breakdown in some cases.  To get around 
this problem, we multiplied the variable {\tt brkdown\_tol} in the 
Bi-CGSTAB algorithm by some small factor, such as $10^{-16}$.
({\tt brkdown\_tol} is normally set equal to {\tt DBL\_EPSILON}, 
which is approximately $2.22\time 10^{-16}$ on our machine).


\section{Problem 1}
\label{prob1}

Problem 1 is taken from Saad \cite[\S 3.7, problem F2DB]{Saad03}.  
It is a two-dimensional PDE
\[ -~ \frac{\partial}{\partial x}(a u_x)
   ~-~ \frac{\partial}{\partial y}(b u_y)
   ~+~ \frac{\partial(du)}{\partial x}
   ~+~ \frac{\partial(eu)}{\partial y}
   ~=~ h,
\]
where 
\[
a(x,y)~=~b(x,y) ~=~ 
\left\{\begin{array}{ll}
	10^3 & \mbox{if~~} \frac{1}{4} < x,~y < \frac{3}{4},\\[5pt]
	1 & \mbox{otherwise}
       \end{array} \right.
\]
and $d(x,y) = 10(x+y)$ and $e(x,y) = 10(x-y)$.  The Dirichlet boundary 
condition $u=0$ is used on the boundary, and the RHS $h$ is immaterial 
since the vector $b$ of the linear system is chosen as $b=Ae$, where 
$A$ is the system matrix and $e=(1,\ldots,1)^T$. 

Table \ref{tbl1} shows the number of iterations and runtimes of the 
various algorithm/preconditioner methods, with and without GS, on a grid 
of $128\time 128$.  In cases of stagnation, the table shows the relative
residual that was achieved before stagnation.  Note that GS enables the 
convergence of Bi-CGSTAB without ILUT; this is useful in a parallel 
setting, because Bi-CGSTAB is inherently parallel but ILUT is not an 
ideal parallel preconditioner.  We can also see that GS is slightly 
helpful to Bi-CGSTAB with ILUT, and, for low-accuracy, also to GMRES 
with and without ILUT.  Another observation is that when GMRES (with 
and without ILUT) stagnates before reaching the prescribed convergence 
goal, GS postpones the stage at which stagnation sets in, and enables 
convergence to a level that is acceptable for most practical 
applications.

\begin{table}[!h]
\def\noconv{\multicolumn{2}{|c|}{no conv.}}
\centering
\begin{tabular}{|l|rl|rl|rl|}
\hline
&\multicolumn{6}{|c|}{}\\[-12pt]
&\multicolumn{6}{|c|}{\bf No.\ of iterations and time (in sec.)}\\
\hline &&&&&&\\[-11pt]
{\bf Method} & \multicolumn{2}{|c|}{rel-res $=\!10^{-4}$}
& \multicolumn{2}{|c|}{rel-res $=\!10^{-7}$}
& \multicolumn{2}{|c|}{rel-res $=\!10^{-10}$} \\
\hline &&&&&&\\[-12pt]
Bi-CGSTAB       & \noconv       & \noconv       & \noconv       \\
with GS         & 91 & (0.30)   & 299 & (0.99)  & 361 & (1.19)  \\
\hline &&&&&&\\[-12pt]
Bi-CGSTAB+ILUT  & 31 & (0.23)   & 107 & (0.67)  & 142 & (0.88)  \\
with GS         & 30 & (0.23)   & 90 & (0.59)   & 130 & (0.81)  \\
\hline 
&\multicolumn{6}{|c|}{}\\[-12pt]
GMRES & \multicolumn{6}{|c|}{converged to $3.8\time 10^{-2}$} \\
with GS & 265 & (0.85) &
\multicolumn{4}{|c|}{converged to $1.1\time 10^{-5}$} \\
\hline 
&\multicolumn{6}{|c|}{}\\[-12pt]
GMRES+ILUT & \multicolumn{6}{|c|}{converged to $3.9\time 10^{-3}$} \\
with GS & 39 & (0.23) &
\multicolumn{4}{|c|}{converged to $1.1\time 10^{-5}$} \\
\hline
\end{tabular}
\caption{No.\ of iterations and runtimes for Problem 1.
         Grid size = $128\time 128$.}
\label{tbl1}
\end{table}

Three additional experiments, based on Problem 1, were also done:
\be
\item The values of $a(x,y)$ and $b(x,y)$ were increased to $10^4$ 
in the inner square.  The results were very similar to those shown
in Table \ref{tbl1}, but with slightly increased runtimes in the
higher-accuracy cases.
\item A ``continuous'' case: the values of $a(x,y)$ and $b(x,y)$ were 
taken as 1 throughout the unit square.  Here, all the methods converged, 
and GS made very little difference.
\item A second continuous case, with $a(x,y)=b(x,y)=1000$.  Table 
\ref{tbl1a} shows the results of this experiment.  As can be seen, 
GS made very little difference.
\ee

\begin{table}[!h]
\centering
\begin{tabular}{|l|rl|rl|rl|}
\hline
&\multicolumn{6}{|c|}{}\\[-12pt]
&\multicolumn{6}{|c|}{\bf No.\ of iterations and time (in sec.)}\\
\hline&&&&&&\\[-11pt]
{\bf Method} & \multicolumn{2}{|c|}{rel-res $=\!10^{-4}$}
& \multicolumn{2}{|c|}{rel-res $=\!10^{-7}$}
& \multicolumn{2}{|c|}{rel-res $=\!10^{-10}$} \\
\hline&&&&&&\\[-12pt]
Bi-CGSTAB        & 121 & (0.41)	& 224 & (0.76)	& 286 & (0.96)  \\
with GS         & 123 & (0.42) & 217 & (0.74) & 261 & (0.88)  \\
\hline&&&&&&\\[-12pt]
Bi-CGSTAB+ILUT   & 39 & (0.28)  & 61 & (0.41) & 85 & (0.56) \\
with GS         & 40 & (0.29)  & 60 & (0.40)  & 85 & (0.56) \\
\hline&&&&&&\\[-12pt]
GMRES           & 1365 & (4.31) & 3704 & (11.69) & 6045 & (19.19) \\
with GS         & 1356 & (4.28) & 3582 & (11.45) & 5722 & (18.16) \\
\hline&&&&&&\\[-12pt]
GMRES+ILUT      & 140 & (0.69)	& 359 & (1.69)	& 579 & (2.67) \\
with GS         & 140 & (0.69) & 359 & (1.69) & 579 & (2.67) \\
\hline
\end{tabular}
\caption{No.\ of iterations and runtimes for a continuous version of
Problem 1, with $a=b=1000$ everywhere.}
\label{tbl1a}
\end{table}

We turn now to studying the effect of GS on the distribution of the 
eigenvalues.  For Problem 1, this was done with a grid size of 
$40\time 40$ for the purpose of a clear presentation.
Table \ref{tbl1b} shows the values of the minimum and maximum eigenvalues 
and the condition number, for the original, the scaled and the continuous 
($a=b=1000$) cases of Problem 1.  Also shown for each case is the number 
of eigenvalues in the first interval (out of 100) in a histogram of the 
eigenvalue distribution.  We can see that while the condition number is 
not changed much by the scaling, the number of eigenvalues in the first 
interval is reduced very significantly.

\begin{table}[!h]
\centering
\begin{tabular}{|l|c|c|c|c|}
\hline&&&&\\[-12pt]
{\bf Matrix} & $\lambda_{\min}$ & $\lambda_{\max}$ &
$\lambda_{\max} / \lambda_{\min}$ &
\parbox[c]{1.25in}{No.\ of eigenvalues in first interval\vspace{2pt}} \\
\hline&&&&\\[-12pt]
Original & 1.87E-2 & 7.96E+3 & 4.25E+5 & 1126 \\
\hline&&&&\\[-12pt]
With GS & 7.07E-6 & 1.78E+0 & 2.52E+5 & 4 \\
\hline&&&&\\[-12pt]
Cont.\ coef. & 1.17E+1 & 8.00E+3 & 6.81E+2 & 8 \\
\hline
\end{tabular}
\caption{Basic eigenvalue information for Problem 1.}
\label{tbl1b}
\end{table}

\begin{figure}[!h]
\begin{tabular}{@{}c@{}c@{}}
\hspace{.1in}
\includegraphics[width=3in]{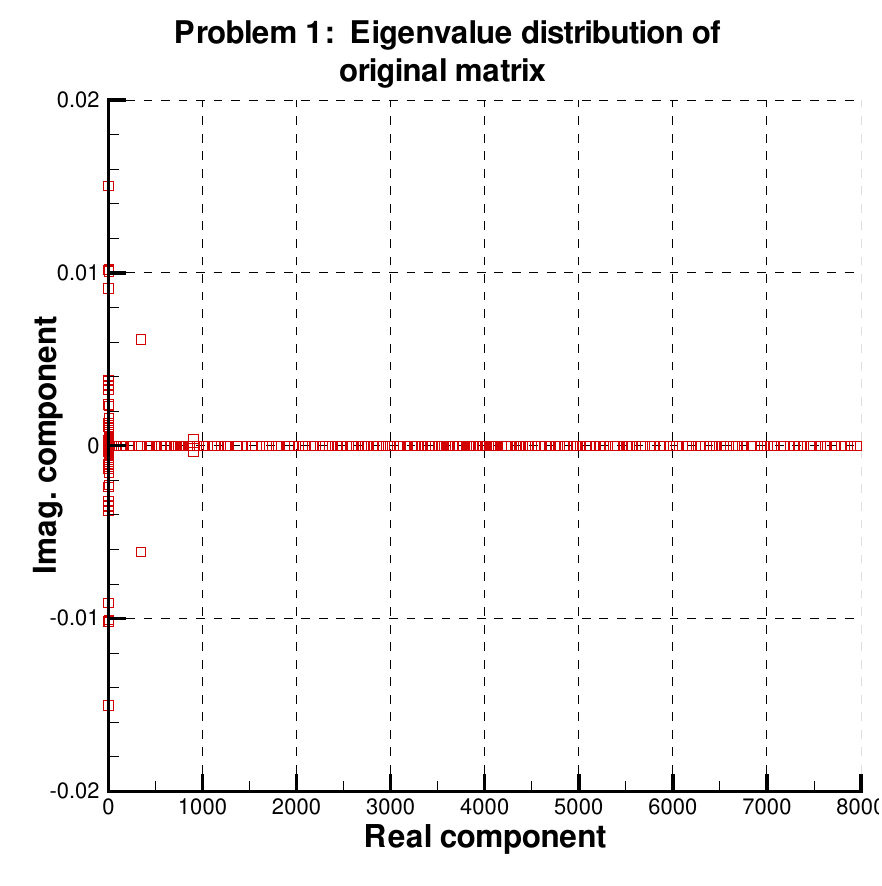}
&
\includegraphics[width=3in]{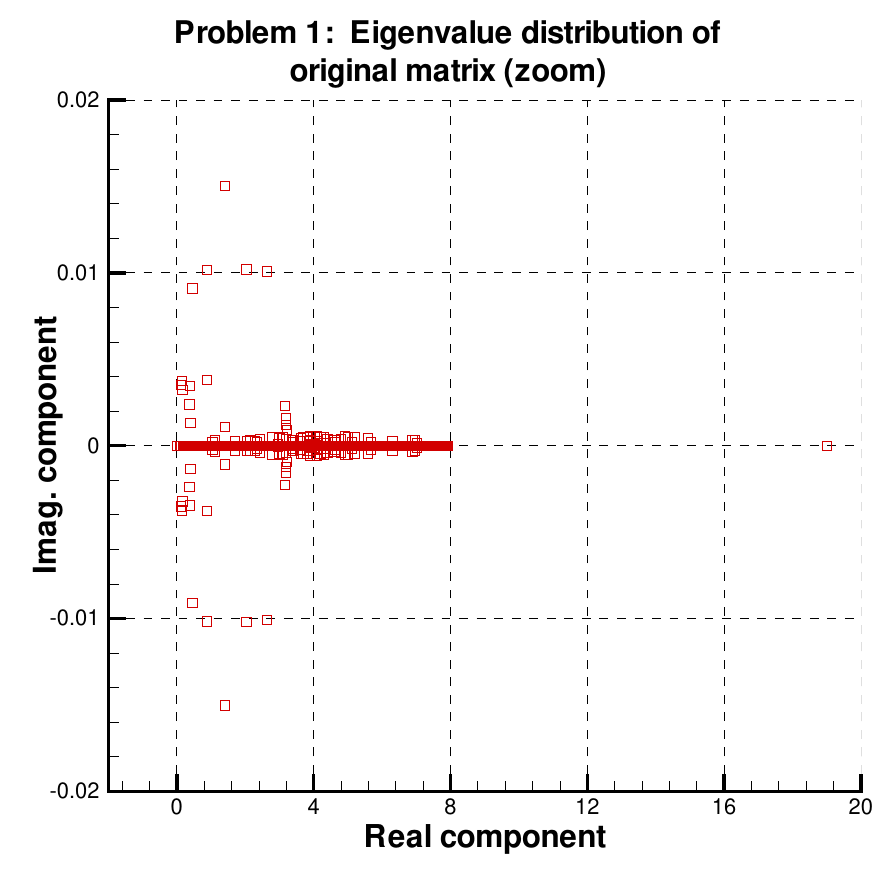}
\end{tabular}
\caption{Eigenvalue distribution of the original matrix of Problem 1,
with a zoom on the range 0--20.}
\label{eigen1}
\end{figure}

\begin{figure}[!h]
\begin{tabular}{@{}c@{}c@{}}
\hspace{.1in}
\includegraphics[width=3in]{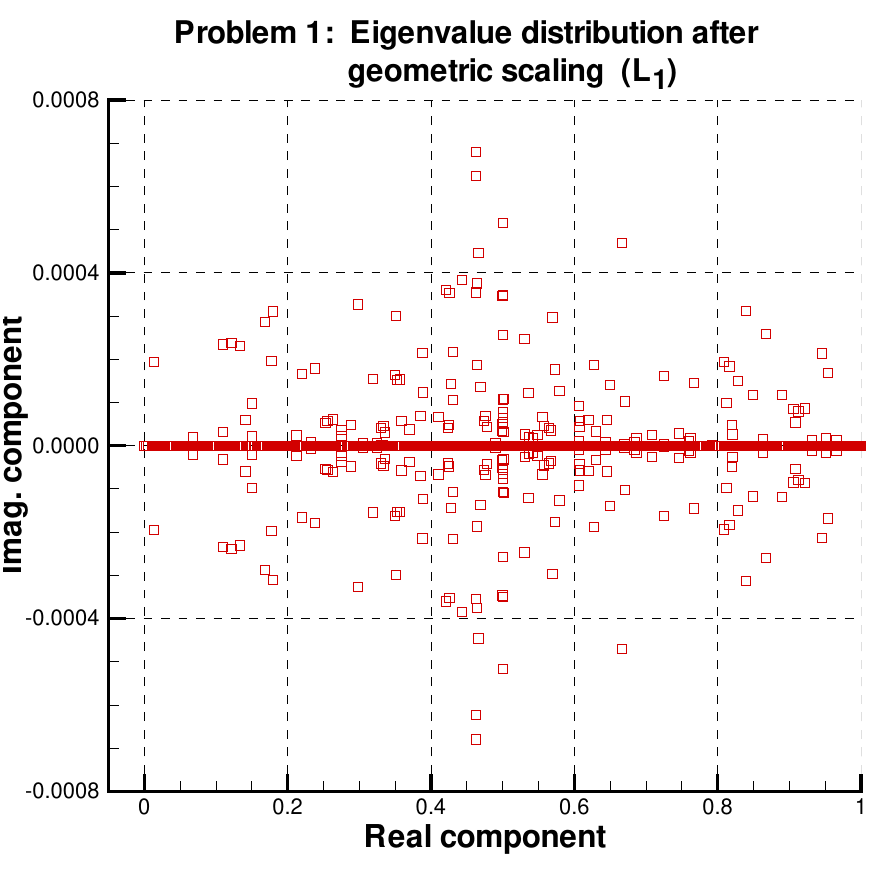}
&
\includegraphics[width=3in]{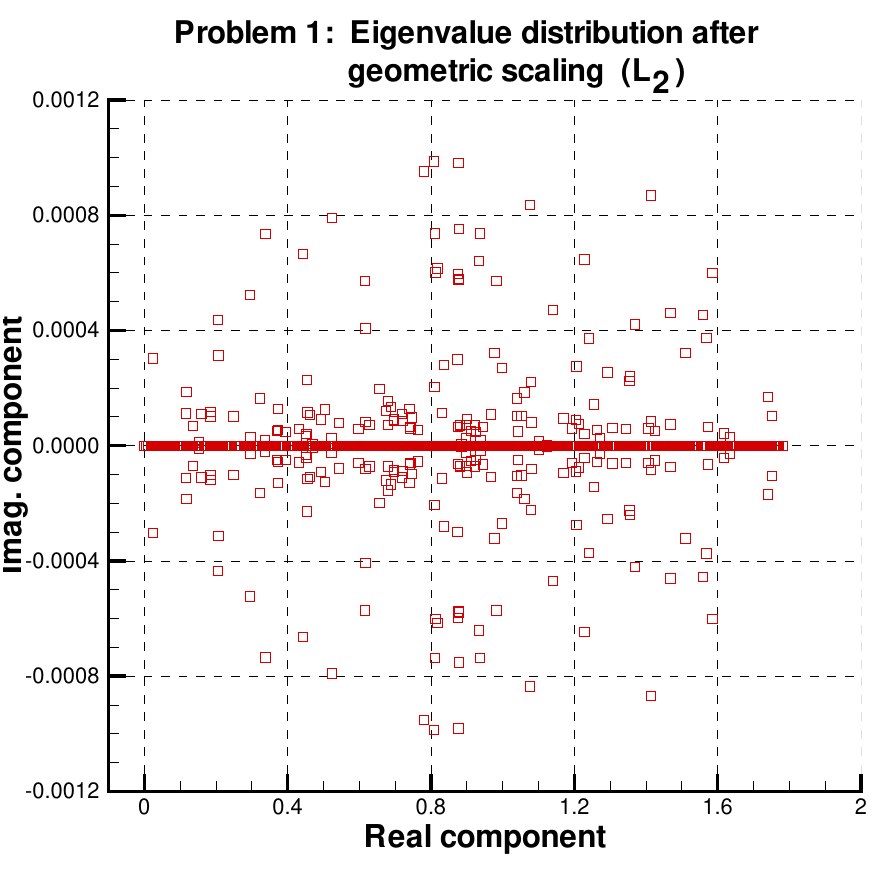}
\end{tabular}
\vspace{-.1in}
\caption{Eigenvalue distribution for Problem 1 after geometric scaling
with the $L_1$- and the $L_2$-norms.}
\label{eigen1a}
\end{figure}

Figure \ref{eigen1} shows the distribution of the eigenvalues of 
Problem 1 at the full scale (left) and with a zoom (right).  We 
can see that there is a congestion of eigenvalues close to zero.
Figure \ref{eigen1a} shows the distribution of the eigenvalues of
Problem 1 after geometric scaling with the $L_1$-norm (left) and the
$L_2$ (right).  Both scalings produce fairly similar distributions;
however we will present results throughout the paper for the $L_2$
scaling.  As to the continuous version of the problem, its eigenvalues
were all real and distributed evenly in the range $(0,8000)$.

\begin{figure}[!h]
\centering
\includegraphics[width=4.5in]{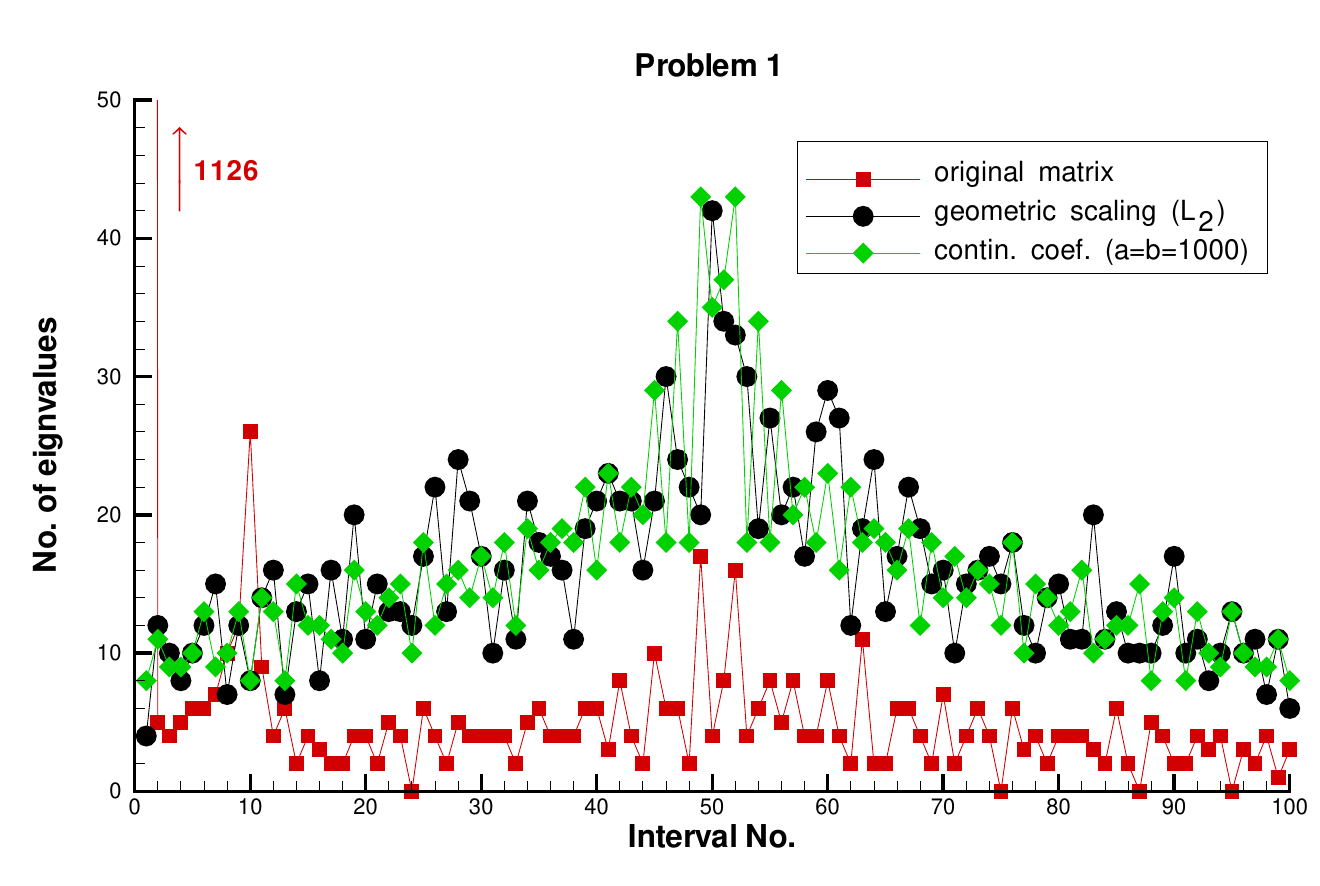}
\vspace{-.1in}
\caption{Histogram of eigenvalues for Problem 1, for the original
matrix, the geometrically scaled matrix, and for a variation of 
Problem 1 with continuous coefficients ($a=b=1000$).}
\label{dist1}
\end{figure}

Figure \ref{dist1} shows a histogram of the eigenvalue distributions 
for the original Problem 1, for the geometrically scaled problem, and 
for the continuous case with $a=b=1000$.

\section{Problem 2}
\label{prob2}

\begin{figure}[!h]
\centering
\includegraphics[width=3in]{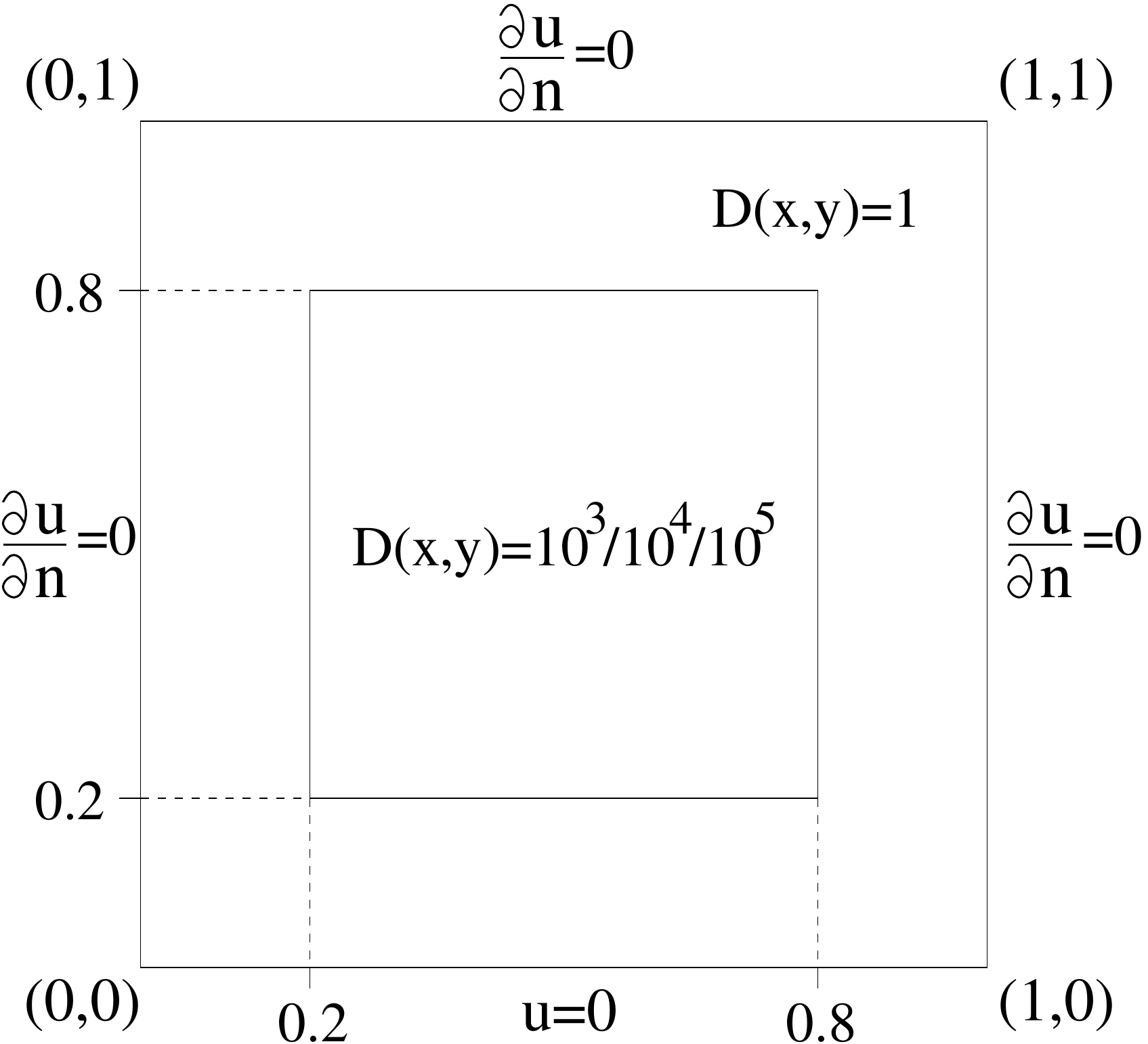}
\caption{Description of Problem 2.}
\label{fig2}
\end{figure}

Problem 2 is based on Example 2 from van der Vorst \cite{Vorst92}, 
to which we added convection terms.  This example demonstrates that 
GS also works with mixed Dirichlet and Neumann boundary conditions.  
The governing PDE is
\[
-~ \frac{\partial(D(x,y)u_x)}{\partial x} ~-~
   \frac{\partial(D(x,y)u_y)}{\partial y} ~+~
   a u_x ~+~ b u_y ~=~ 1,
\label{eq2}
\]
with $D(x,y)$ and boundary conditions as shown in Figure \ref{fig2}.  
The convection terms were $a=b=200$.
In the original example, the internal value of $D$ is $10^3$, but
we also tested internal values of $10^4$ and $10^5$.  Also, the 
internal square in our case is somewhat smaller.
The unit square was discretized with a grid size of $150\time 150$.
Together with the boundary equations, the system consisted of 22,952
equations.  The resulting system is indefinite, with eigenvalues in 
the four quadrants of the imaginary plane.

Tables \ref{tbl2}, \ref{tbl2a} and \ref{tbl2b} show the number of 
iterations and runtimes of the various algorithm and preconditioner 
combinations, with and without GS, for $D=10^3$, $D=10^4$, and
$D=10^5$, respectively.  When there was no convergence to the 
prescribed goal, the table shows the relative residual that was 
achieved.  

\begin{table}[!h]
\centering
\begin{tabular}{|l|rl|rl|rl|}
\hline
&\multicolumn{6}{|c|}{}\\[-12pt]
&\multicolumn{6}{|c|}{\bf No.\ of iterations and time (in sec.)}\\
\hline&&&&&&\\[-11pt]
{\bf Method} & \multicolumn{2}{|c|}{rel-res $=\!10^{-4}$}
& \multicolumn{2}{|c|}{rel-res $=\!10^{-7}$}
& \multicolumn{2}{|c|}{rel-res $=\!10^{-10}$} \\
\hline&&&&&&\\[-12pt]
Bi-CGSTAB       & \noconv      & \noconv       & \noconv       \\
with GS         & 591 & (2.13) & 1025 & (3.69) & 
\multicolumn{2}{|c|}{conv.\ to $7.23\time10^{-10}$} \\
\hline&&&&&&\\[-12pt]
Bi-CGSTAB+ILUT  & 92 & (0.75)  & 139 & (1.09)  & 215 & (1.68)  \\
with GS         & 88 & (0.72)  & 125 & (1.00)  & 324 & (2.45)  \\
\hline
&\multicolumn{6}{|c|}{}\\[-12pt]
GMRES & \multicolumn{6}{|c|}{converged to 6.15}  \\
with GS & \multicolumn{6}{|c|}{converged to $2.145\time10^{-3}$} \\
\hline
& \multicolumn{6}{|c|}{}\\[-12pt]
GMRES+ILUT & \multicolumn{6}{|c|}{converged to 0.927} \\
with GS & 3361 & (19.86) & 
\multicolumn{4}{|c|}{converged to $3.67\time10^{-5}$} \\
\hline
\end{tabular}
\caption{No.\ of iterations and runtimes for Problem 2 
(internal $D=10^3$).}
\label{tbl2}
\end{table}

\begin{table}[!h]
\centering
\begin{tabular}{|l|rl|rl|rl|}
\hline
&\multicolumn{6}{|c|}{}\\[-12pt]
&\multicolumn{6}{|c|}{\bf No.\ of iterations and time (in sec.)}\\
\hline&&&&&&\\[-11pt]
{\bf Method} & \multicolumn{2}{|c|}{rel-res $=\!10^{-4}$}
& \multicolumn{2}{|c|}{rel-res $=\!10^{-7}$}
& \multicolumn{2}{|c|}{rel-res $=\!10^{-10}$} \\
\hline&&&&&&\\[-12pt]
Bi-CGSTAB       & \noconv      & \noconv      & \noconv      \\
with GS         & 589 & (2.01) & 707 & (2.41) & 1493 & (5.37) \\
\hline&&&&&&\\[-12pt]
Bi-CGSTAB+ILUT  & 96  & (0.76) & 192 & (1.44) & 255 & (1.86) \\
with GS         & 79  & (0.64) & 121 & (0.92) & 163 & (1.19) \\
\hline
&\multicolumn{6}{|c|}{}\\[-12pt]
GMRES & \multicolumn{6}{|c|}{converged to 1.84} \\
with GS & 
\multicolumn{6}{|c|}{converged to $2.19\time10^{-4}$} \\
\hline
&\multicolumn{6}{|c|}{}\\[-12pt]
GMRES+ILUT & \multicolumn{6}{|c|}{converged to 0.924} \\
with GS    & 
\multicolumn{6}{|c|}{converged to $1.90\time10^{-4}$} \\
\hline
\end{tabular}
\caption{No.\ of iterations and runtimes for Problem 2
(internal $D=10^4$).}
\label{tbl2a}
\end{table}

\begin{table}[!h]
\centering
\begin{tabular}{|l|rl|rl|rl|}
\hline
&\multicolumn{6}{|c|}{}\\[-12pt]
&\multicolumn{6}{|c|}{\bf No.\ of iterations and time (in sec.)}\\
\hline&&&&&&\\[-11pt]
{\bf Method} & \multicolumn{2}{|c|}{rel-res $=\!10^{-4}$}
& \multicolumn{2}{|c|}{rel-res $=\!10^{-7}$}
& \multicolumn{2}{|c|}{rel-res $=\!10^{-10}$} \\
\hline&&&&&&\\[-12pt]
Bi-CGSTAB       & \noconv      & \noconv      & \noconv      \\
with GS         & 224 & (0.82) & 730 & (2.65) & 
\multicolumn{2}{|c|}{conv.\ to $1.69\time10^{-9}$} \\
\hline&&&&&&\\[-12pt]
Bi-CGSTAB+ILUT  & 96  & (1.15) & 192 & (1.56) & 
\multicolumn{2}{|c|}{conv.\ to $5.16\time10^{-10}$} \\
with GS         & 21  & (0.23) & 125 & (0.98) & 217 & (1.68) \\
\hline
&\multicolumn{6}{|c|}{}\\[-12pt]
GMRES & \multicolumn{6}{|c|}{converged to 1.37} \\
with GS & 357 & (1.38) & 
\multicolumn{4}{|c|}{converged to $2.19\time10^{-5}$} \\
\hline
&\multicolumn{6}{|c|}{}\\[-12pt]
GMRES+ILUT & \multicolumn{6}{|c|}{converged to 0.90} \\
with GS    & 40 & (0.32) & 
\multicolumn{4}{|c|}{converged to $1.89\time10^{-5}$} \\
\hline
\end{tabular}
\caption{No.\ of iterations and runtimes for Problem 2 
(internal $D=10^5$).}
\label{tbl2b}
\end{table}

The tables show that GS was helpful in all cases, with the
exception of Bi-CGSTAB with ILUT in Table \ref{tbl2}, where
it increased the number of iterations required in the highest
accuracy case.  A close examination of this case revealed that
starting from iteration 220, BiCGSTAB with ILUT (after GS) was
extremely oscillatory, with fluctuations of up to three orders
of magnitude in the relative residual; the required convergence 
goal of $10^{-10}$ was almost attained at 225 iterations.

The three tables show that the behavior of GS is not necessarily 
``smooth'': consider the results of Bi-CGSTAB with ILUT for 
rel-res = $10^{-10}$: we can see that the goal was not reached 
for $D=10^3$ and $D=10^5$, but it was reached for $D=10^4$.  The 
explanation for this is the large number of iterations required
by Bi-CGSTAB in this case, as seen in Table \ref{tbl2a} and the
oscillatory nature of Bi-CGSTAB.  The accumulated roundoff error 
is sometimes too large to reach the convergence goal, and while 
the updated error measured by the algorithm showed convergence, 
the true relative residual did not always decrease sufficiently.

Another interesting point is the behavior of GS with GMRES and ILUT 
in the low-accuracy case:  for $D=10^3$, there was convergence after
many iterations, for $D=10^4$ there was no convergence, and then for
$D=10^5$ there was a very fast convergence.  In order to examine this
behavior, we also tested this particular case with $D=10^6$, and we
noticed that starting $D=10^4$, GS improved the convergence of GMRES
by one order of magnitude for every increase in $D$ by one order of 
magnitude.  Not doubt, this phenomenon should be studied further.
In any case, for the larger values of $D$, GS enabled the convergence
of GMRES, with and without ILUT, to practical levels of accuracy.

For the eigenvalue data, we discretized the domain by a grid 
of $40\time40\time40$.  Table \ref{tbl2d} provides the basic 
eigenvalue information for Problem 2, for the original ($D=10^3$) 
and the scaled matrices.  Also shown are the eigenvalues for a 
variation of Problem 2 with ``continuous'' coefficients, obtained 
by taking $D=10^3$ throughout the unit square.  The last column of 
the table shows the number of eigenvalues whose real part lies in 
the same interval as the origin, when the range of (the real parts) 
of the eigenvalues is divided into 100 intervals.  We can see that 
GS reduces the condition number by three orders of magnitude, and
the reduction is even greater w.r.t.\ the continuous case.
Furthermore, the number of eigenvalues around the origin is reduced 
very significantly, even below that of the continuous case.

\begin{table}[!h]
\centering
\begin{tabular}{|l|c|c|c|c|}
\hline&&&&\\[-11pt]
{\bf Matrix} & $\lambda_{\min}$ & $\lambda_{\max}$ &
$\lambda_{\max} / \lambda_{\min}$ &
\parbox[c]{.8in}{Eigenvalues around $x\!=\!0$\vspace{3.5pt}} \\
\hline&&&&\\[-12pt]
Original $(D\!=\!10^3)$ & 3.33E-1 & 1.34E+7 & 4.02E+7 & 1058 \\
\hline&&&&\\[-12pt]
With GS      & 9.15E-5 & 1.78E0 & 1.95E+4 & 8 \\
\hline&&&&\\[-12pt]
Continuous coef.\ & 1.48E-2 & 1.34E+7 & 9.09E+8 & 130 \\
\hline
\end{tabular}
\caption{Basic eigenvalue information for Problem 2.}
\label{tbl2d}
\end{table}

\begin{figure}[!h]
\begin{tabular}{@{}c@{}c@{}}
\hspace{.1in}
\includegraphics[width=3in]{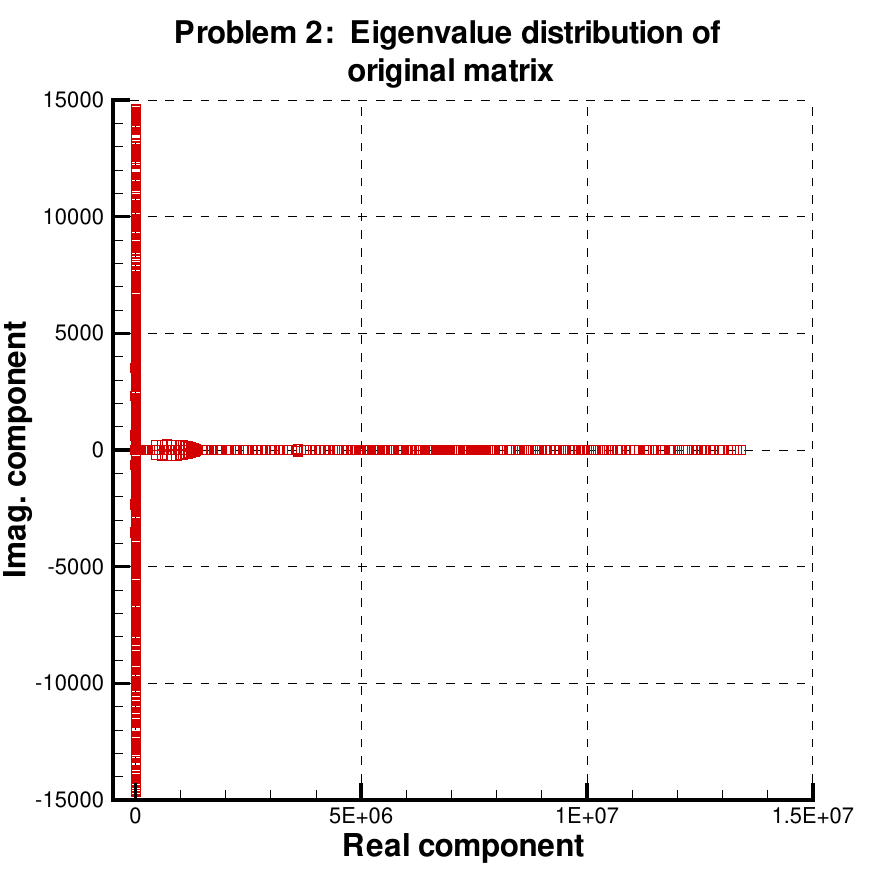}
&
\includegraphics[width=3in]{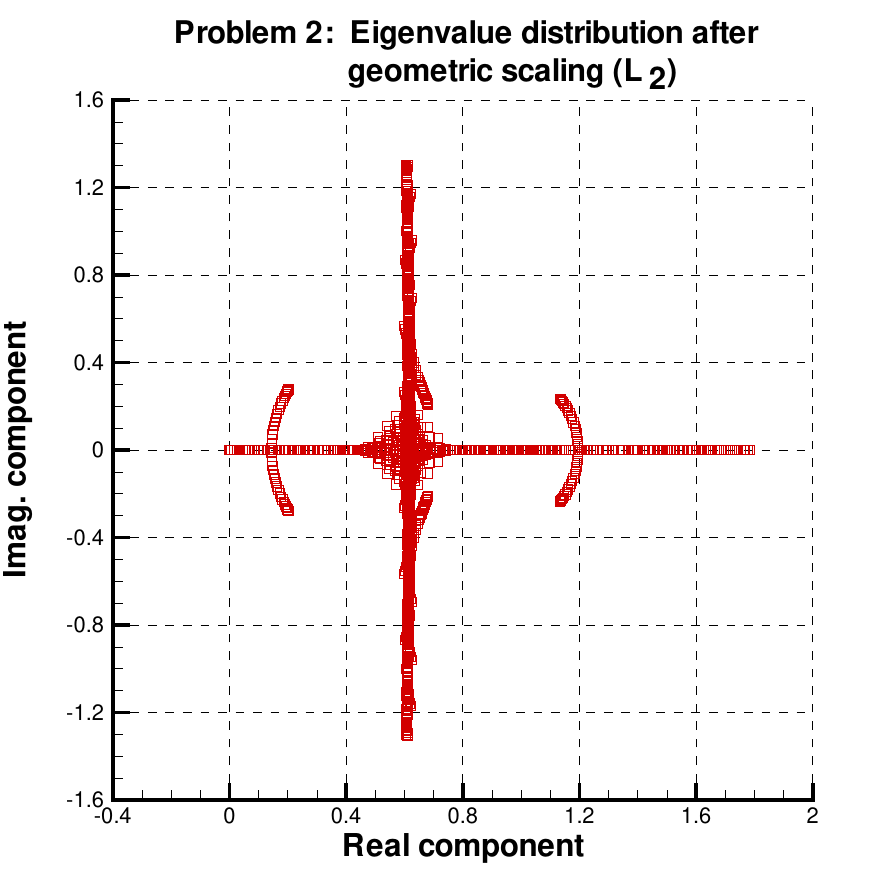} 
\end{tabular}
\caption{Eigenvalue distribution for Problem 2, for the original
and  the scaled
cases.}
\label{eigen2}
\end{figure}

Figure \ref{eigen2} shows the distribution of the eigenvalues for 
Problem 2 with $D=10^3$, and for the geometrically scaled problem. 
It can be seen that the eigenvalues of the original matrix are very 
concentrated around the origin.  In the scaled matrix, there are
very few eigenvalues around the origin, but many eigenvalues have
their real part around 0.6.

\begin{figure}[!h]
\centering
\includegraphics[width=4.5in]{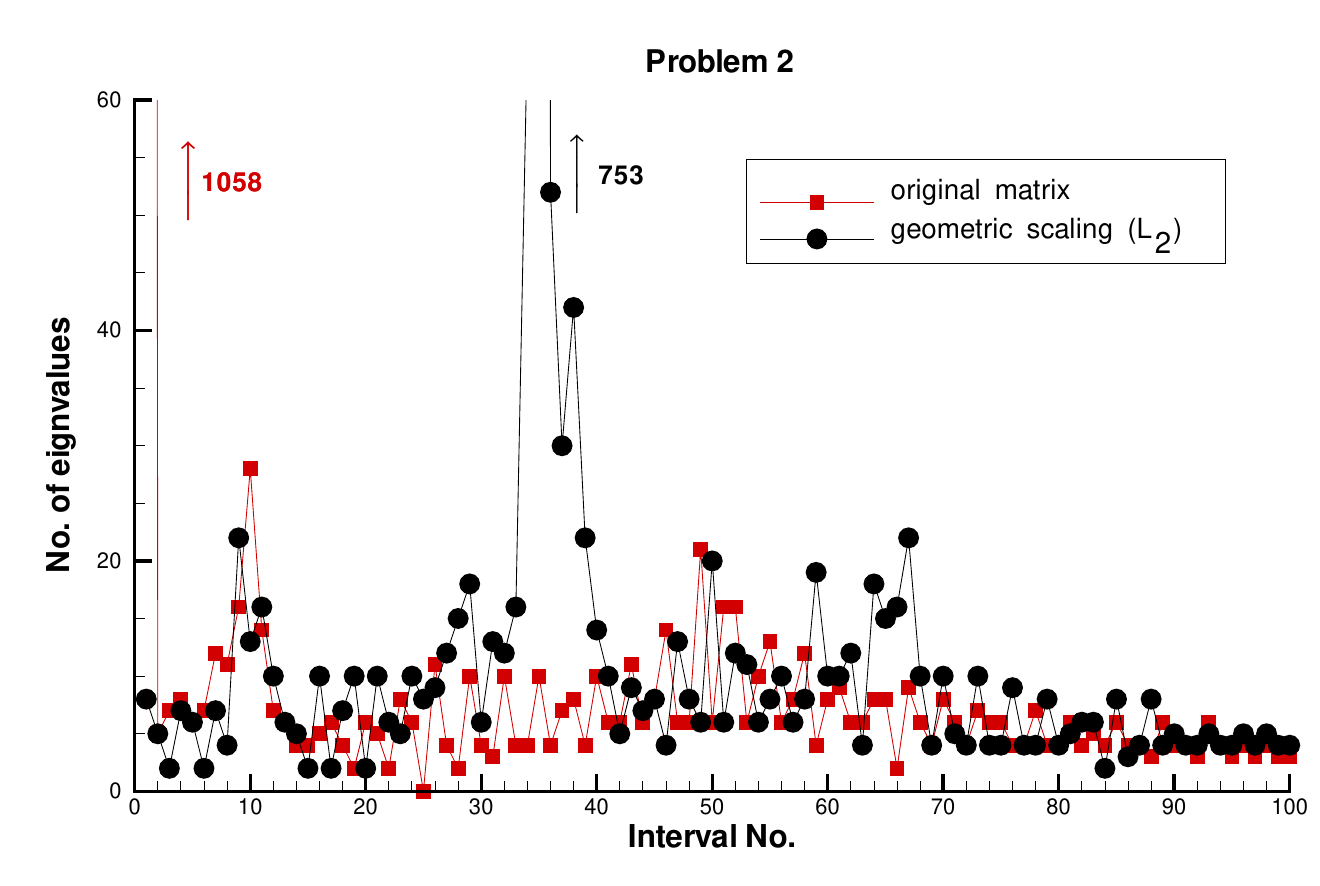}
\vspace{-.1in}
\caption{Histogram of the real part of the eigenvalues for Problem 2, 
for the original and the scaled cases.}
\label{dist2}
\end{figure}

Figure \ref{dist2} provides another view of the eigenvalue distribution 
with a histogram of the real part of the eigenvalues for the original 
and the scaled matrices.  

\section{Problem 3}
\label{prob3}

\begin{figure}[!h]
\centering
\includegraphics[width=3in]{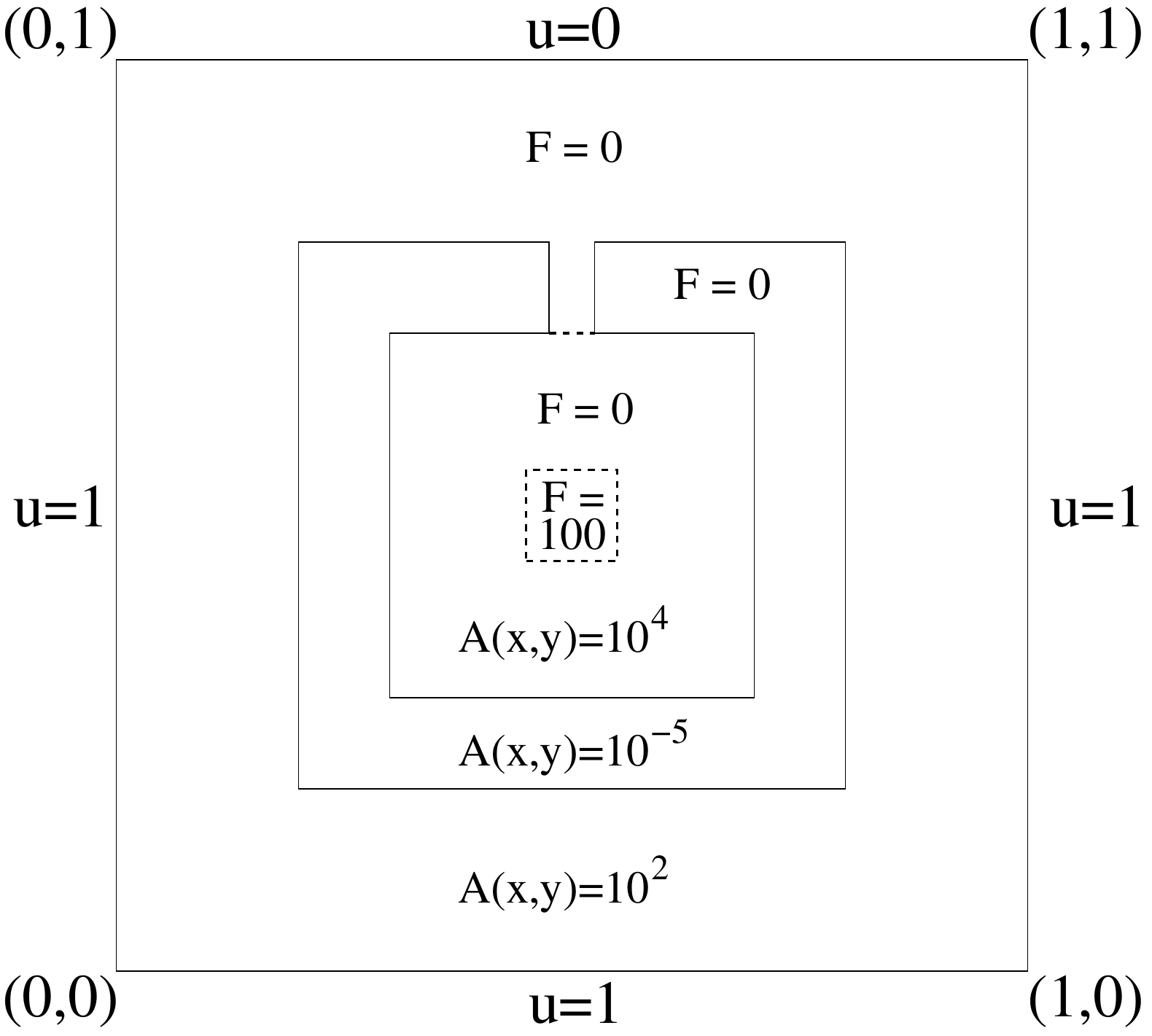}
\caption{Description of Problem 3.}
\label{fig3}
\end{figure}

Problem 3 is also taken from van der Vorst \cite[Example 4]{Vorst92}.
It describes a certain groundwater flow problem which leads to a 
nonsymmetric system, with a complex geometry and several jumps in 
the discontinuities of the equations.  This problem is well-known
for its difficulty.  The basic equation is the following:
\[
 -~ \frac{\partial}{\partial x} \left(A(x,y)u_x\right) 
~-~ \frac{\partial}{\partial y} \left(A(x,y)u_y\right) 
~+~ B(x,y)u_x ~=~ F,
\]
where $A(x,y)$ and $F$ are taken as shown in Figure \ref{fig3}, and
$B(x,y) = 2\exp(2(x^2+y^2))$.  The Dirichlet boundary conditions are 
taken as shown in Figure \ref{fig3}.
The unit square was discretized with a grid size of $128\time 128$,
resulting in 16,129 equations. 

Nothing converged without ILUT, so we present in Table \ref{tbl3} 
only the results with ILUT, with and without GS.  On this difficult 
problem, GS reduced the runtime of Bi-CGSTAB with ILUT by about 25\%.  
GS also enabled the convergence of GMRES with ILUT, though the 
achieved runtimes were an order of magnitude larger than those 
of the scaled Bi-CGSTAB with ILUT in the higher-accuracy cases.

\begin{table}[!h]
\centering
\begin{tabular}{|l|rl|rl|rl|}
\hline
&\multicolumn{6}{|c|}{}\\[-12pt]
&\multicolumn{6}{|c|}{\bf No.\ of iterations and time (in sec.)}\\
\hline&&&&&&\\[-11pt]
{\bf Method} & \multicolumn{2}{|c|}{rel-res $=\!10^{-4}$}
& \multicolumn{2}{|c|}{rel-res $=\!10^{-7}$}
& \multicolumn{2}{|c|}{rel-res $=\!10^{-10}$} \\
\hline&&&&&&\\[-12pt]
Bi-CGSTAB+ILUT   & 93  & (0.60)  & 124 & (0.79)  & 152 & (0.95)  \\
with GS         & 67  & (0.44)  & 90  & (0.59)  & 112 & (0.72)  \\
\hline&&&&&&\\[-12pt]
GMRES+ILUT      & \noconv       & \noconv       & \noconv       \\
with GS         & 338 & (1.58)  & 1008 & (4.61) & 1683 & (7.71) \\
\hline
\end{tabular}
\caption{No.\ of iterations and runtimes for Problem 3.}
\label{tbl3}
\end{table}

For the eigenvalue data, we discretized Problem 3 with a grid of 
$40\time 40$.  The data for the original and the scaled matrices
is summarized in Table \ref{tbl3a}, which also shows the number 
of eigenvalues in the first interval (out of 100) of the eigenvalue 
histogram.  We can see that the condition number is decreased very 
significantly, and so is the number of eigenvalues in the first 
interval.

\begin{table}[!h]
\centering
\begin{tabular}{|l|c|c|c|c|}
\hline&&&&\\[-12pt]
{\bf Matrix} & $\lambda_{\min}$ & $\lambda_{\max}$ &
$\lambda_{\max} / \lambda_{\min}$ &
\parbox[c]{1.25in}{No.\ of eigenvalues in first interval\vspace{2pt}} \\
\hline&&&&\\[-12pt]
Original        & 6.97E-5 & 7.93E+4 & 1.14E+9 & 1284 \\
\hline&&&&\\[-12pt]
With GS      & 1.21E-4 & 1.78E+0 & 1.47E+4  & 167 \\
\hline
\end{tabular}
\caption{Basic eigenvalue information for Problem 3.}
\label{tbl3a}
\end{table}

\begin{figure}[!h]
\begin{tabular}{@{}c@{}c@{}}
\hspace{.1in}
\includegraphics[width=3in]{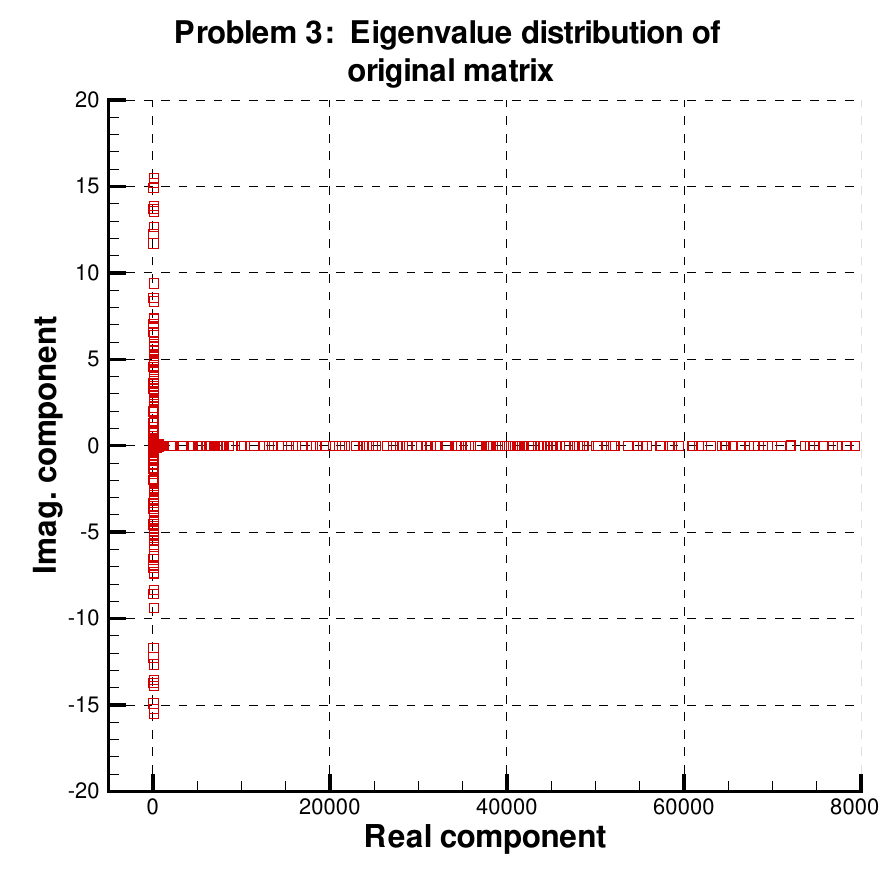}
&
\includegraphics[width=3in]{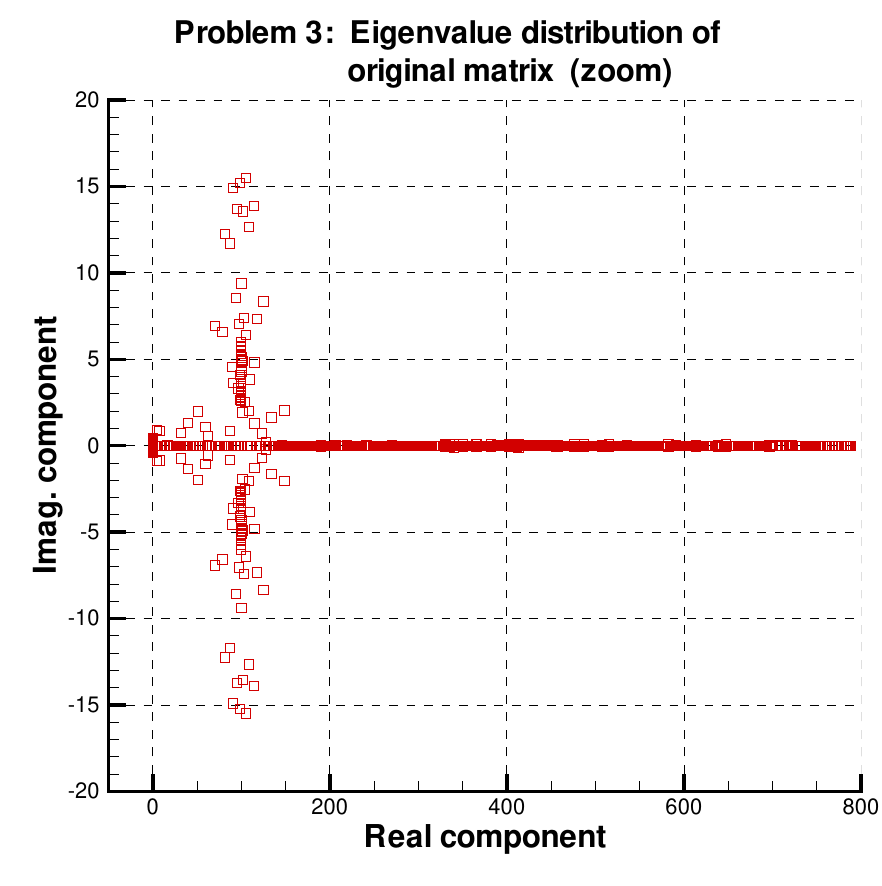} \\
\multicolumn{2}{c}{\includegraphics[width=3in]{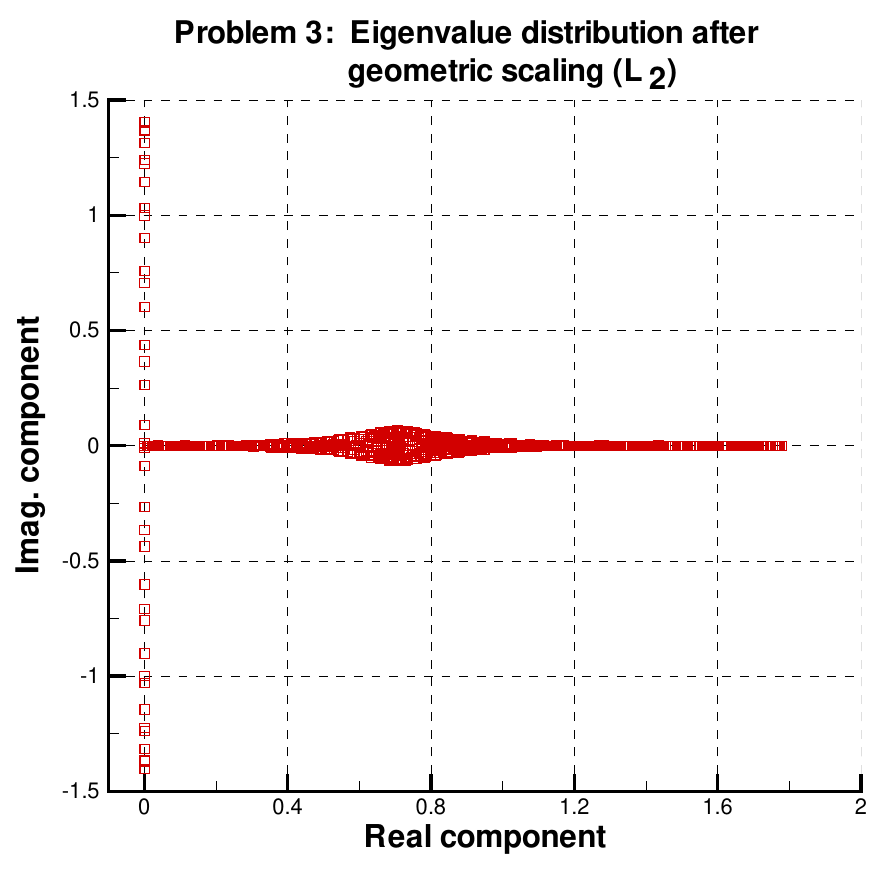}} \\
\end{tabular}
\caption{Eigenvalue distribution for Problem 3, with a zoom to 
the region 0--800, and the distribution for the scaled matrix.} 
\label{eigen3}
\end{figure}

Figure \ref{eigen3} shows the distribution of the eigenvalues 
for the original (with a zoom) and the scaled matrices of 
Problem 3.

\begin{figure}[!h]
\centering
\includegraphics[width=4.5in]{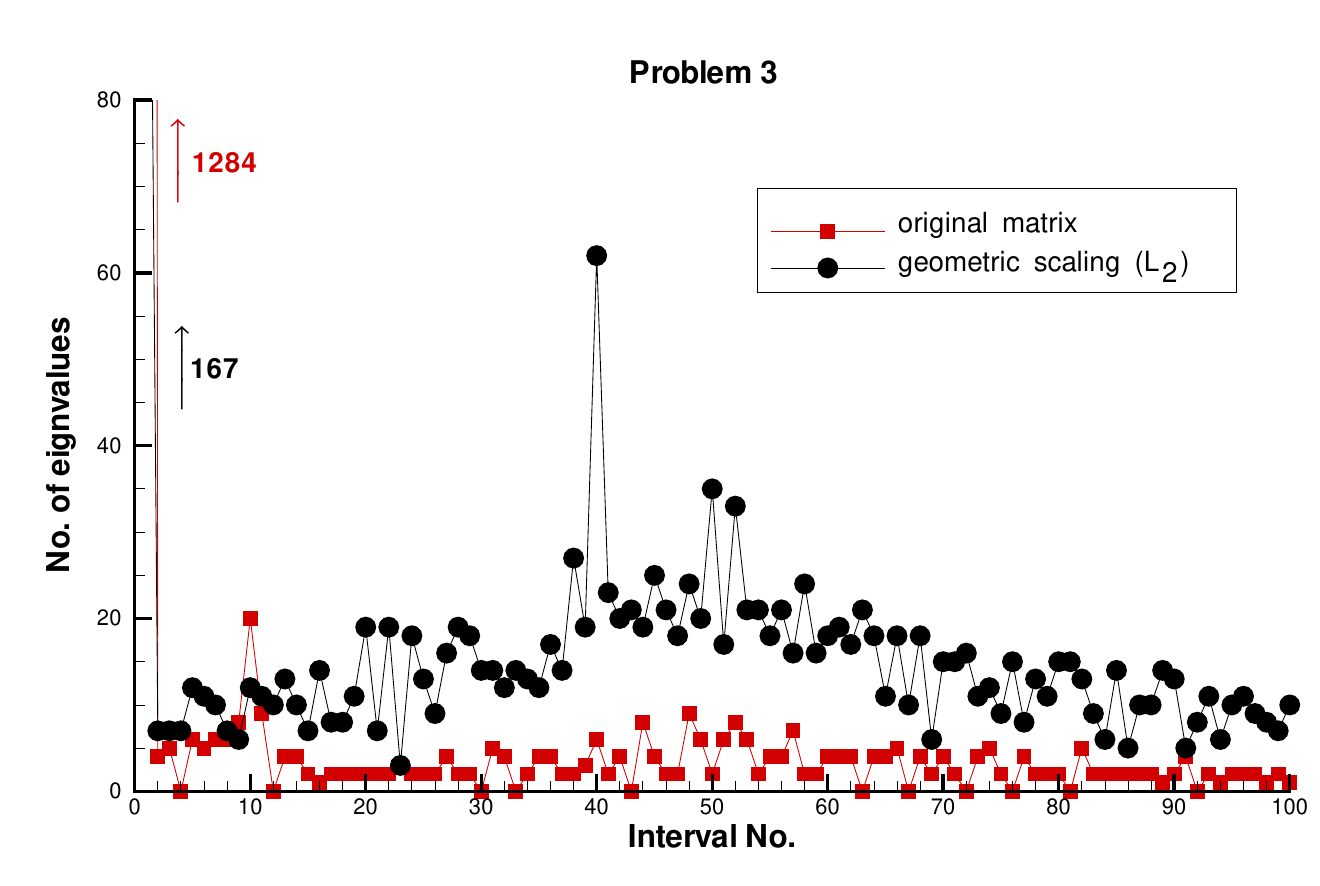}
\vspace{-.1in}
\caption{Histogram of the eigenvalues for Problem 3, for the original
and the scaled cases.}
\label{dist3}
\end{figure}

Figure \ref{dist3} is a histogram of the eigenvalues of the 
original and the scaled matrices.  We can see from Figures 
\ref{eigen3} and \ref{dist3} that in the original matrix, 
the eigenvalues are very concentrated around the origin, 
while in the scaled matrix, many are ``pushed'' away from 
the origin.

\section{Problem 4}
\label{prob4}

This problem is based on a three-dimensional problem from Graham 
and Hagger \cite{Graham99}, to which we added convection terms.  
The differential equation is the following:
\[
 -~ \frac{\partial}{\partial x}(a u_x)
~-~ \frac{\partial}{\partial y}(a u_y) 
~-~ \frac{\partial}{\partial z}(a u_z) 
~+~ du_x ~+~ eu_y ~+~ fu_z ~=~ 0,
\]
where the domain is the unit cube, and $a(x,y,z)$ is defined as 
\[
a(x,y,z) ~=~ 
	 \left\{ \begin{array}{ll}
	 D & \mbox{if~~} \frac{1}{3} < x,~y,~z < \frac{2}{3}, \\[2pt]
	 1 & \mbox{otherwise.} 
		 \end{array} \right.
\]
The Dirichlet boundary conditions are prescribed with $u=1$ on the 
$z=0$ plane and $u=0$ on the other boundaries.  The convection terms 
were taken as $d=e=f=100$, and two values of $D$ were tested:
$D=10^4$ (as in the original problem) and $D=10^6$.  Two grids were
tested: $40\time40\time40$ and $80\time80\time80$.  The resulting 
linear systems are indefinite, with eigenvalues in the four quadrants 
of the imaginary plane.  This problem will also be used to test the
limit of usefulness of GS as the convection increases.

Due to the many different cases, the data for this problem is 
presented differently.
Table \ref{tbl4} shows the number of iterations required for 
convergence, for a grid of $80\time80\time80$, with $D=10^4$ 
and $D=10^6$.  We can see that GS enabled the convergence of 
Bi-CGSTAB without ILUT, and it speeded up Bi-CGSTAB with ILUT 
quite significantly.  GS also enabled the convergence of GMRES, 
with and withou ILUT, for rel-res = $10^{-4}$, and also for 
rel-res = $10^{-7}$ with $D=10^6$.

\begin{table}[!h]
\centering
\begin{tabular}{|l|c|c|c|c|c|c|}
\hline
& \multicolumn{2}{|c|}{}
& \multicolumn{2}{|c|}{}
& \multicolumn{2}{|c|}{}
\\[-11pt]
& \multicolumn{2}{|c|}{rel-res $=\!10^{-4}$}
& \multicolumn{2}{|c|}{rel-res $=\!10^{-7}$}
& \multicolumn{2}{|c|}{rel-res $=\!10^{-10}$} \\
\hline&&&&&&\\[-11pt]
{\bf Method} & $D\!=\!10^4$ & $D\!=\!10^6$ & $D\!=\!10^4$ & 
$D=10^6$ & $D=10^4$ & $D=10^6$ \\
\hline
&&&&&&\\[-12pt]
Bi-CGSTAB &---&---&---&---&---&--- \\
with GS        & 112 & 111 & 262 & 160 & 406 & 374 \\
\hline&&&&&&\\[-12pt]
Bi-CGSTAB+ILUT & 77 & 52 & 112 & 139 & 123 & 169 \\
with GS         & 13 & 13 & 52  & 19 & 84  & 86 \\
\hline
&&&&&&\\[-12pt]
GMRES &---&---&---&---&---&--- \\
with GS & 208 & 207 &---& 286 &---&--- \\
\hline
&&&&&&\\[-12pt]
GMRES+ILUT &---&---&---&---&---&--- \\
with GS & 28 & 28 &---& 46 &---&--- \\
\hline
\multicolumn{7}{|l|}{} \\[-12pt]
\multicolumn{7}{|l|}{Note: ``---'' denotes no convergence 
to the prescribed accuracy.} \\
\hline
\end{tabular}
\caption{No.\ of iterations for Problem 4, with $D=10^4$, $D=10^6$,
and grid $80\time 80\time 80$.}
\label{tbl4}
\end{table}

\begin{figure}[!h]
\hspace{5pt}
\begin{tabular}{@{}c@{}c@{}}
Results for $D = 10^4$ & Results for $D = 10^6$
\\ \hline
&\\[-18pt]
\includegraphics[width=3.15in]{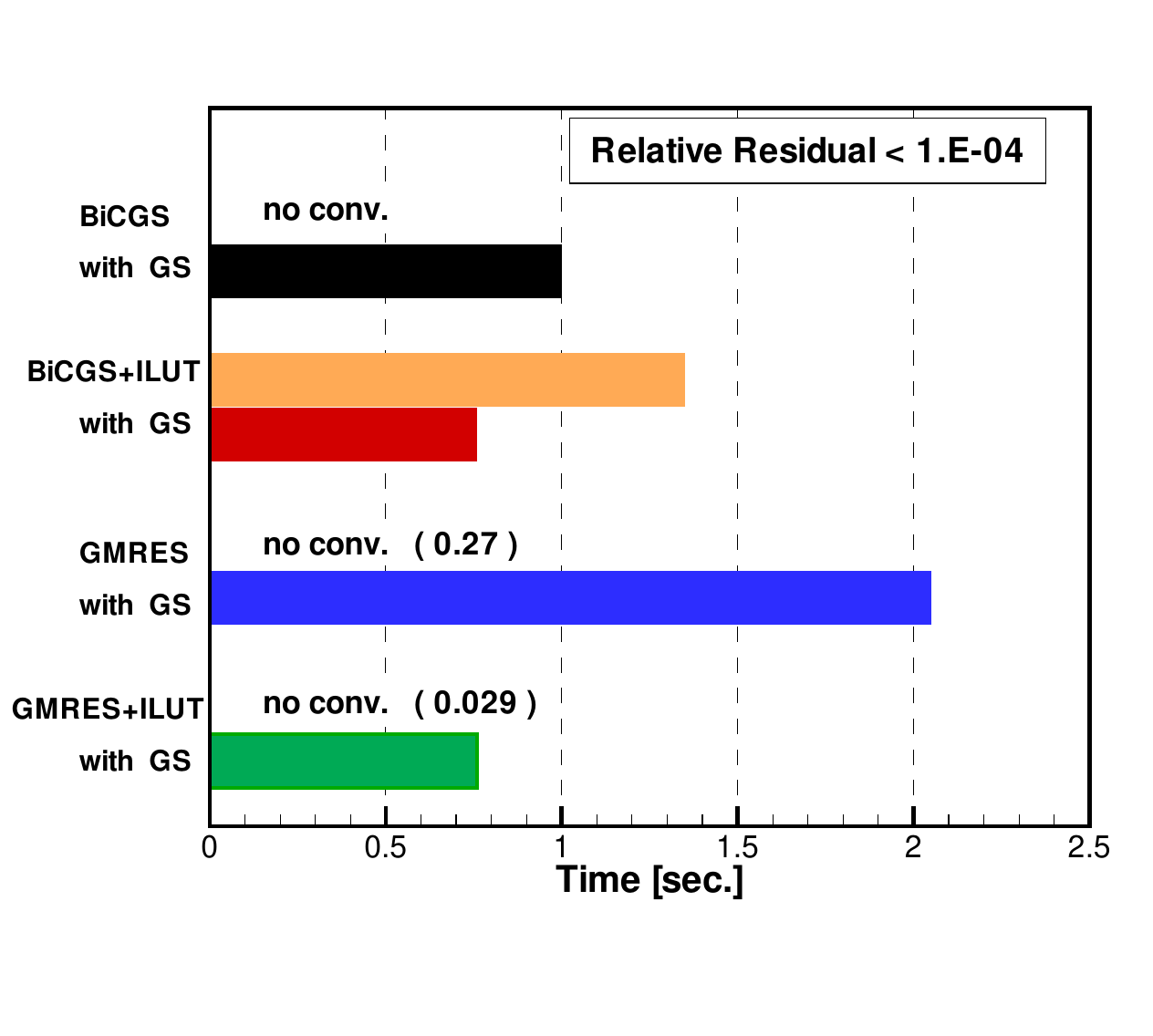}
&
\includegraphics[width=3.15in]{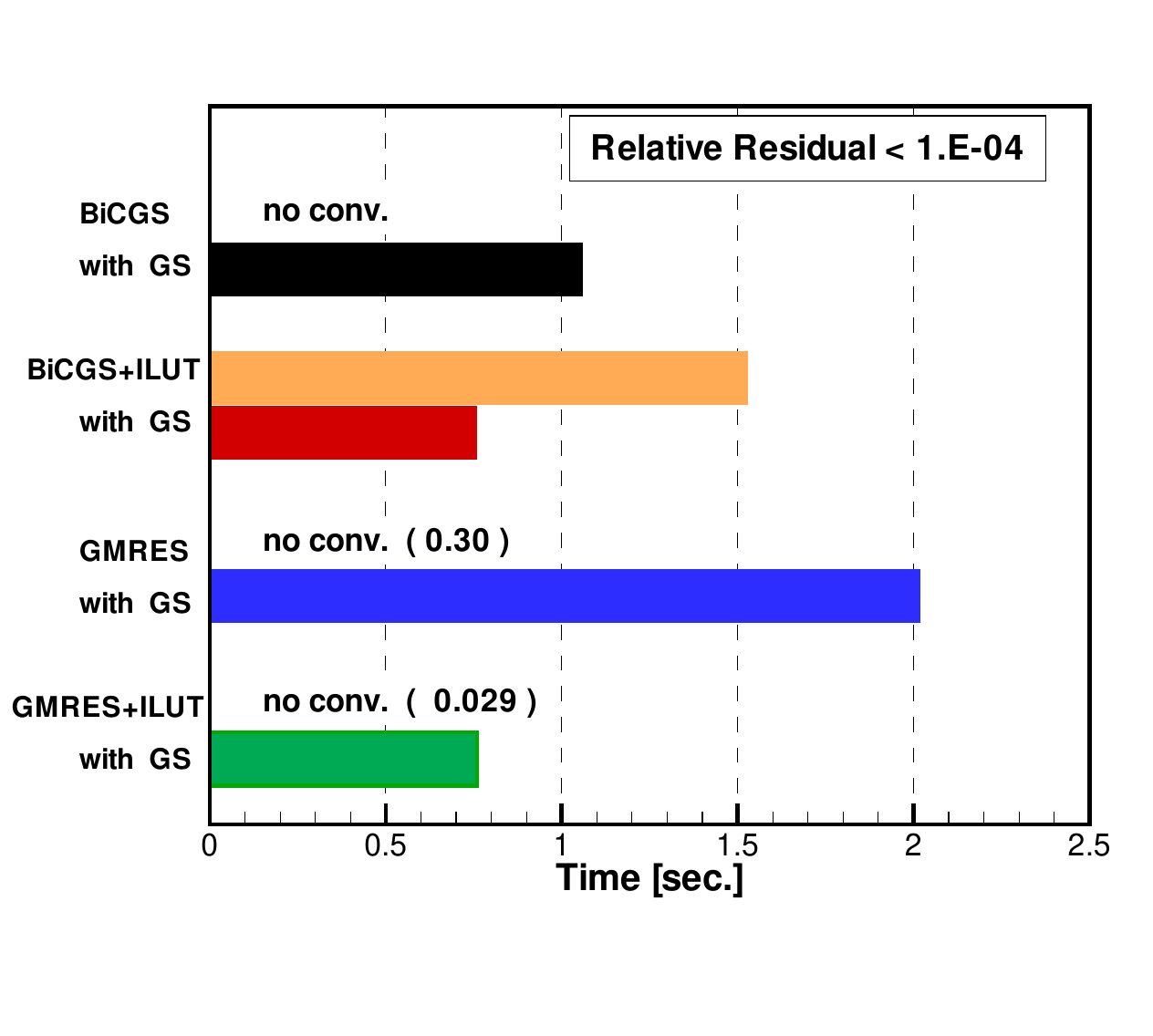}
\\[-.5in]
\includegraphics[width=3.15in]{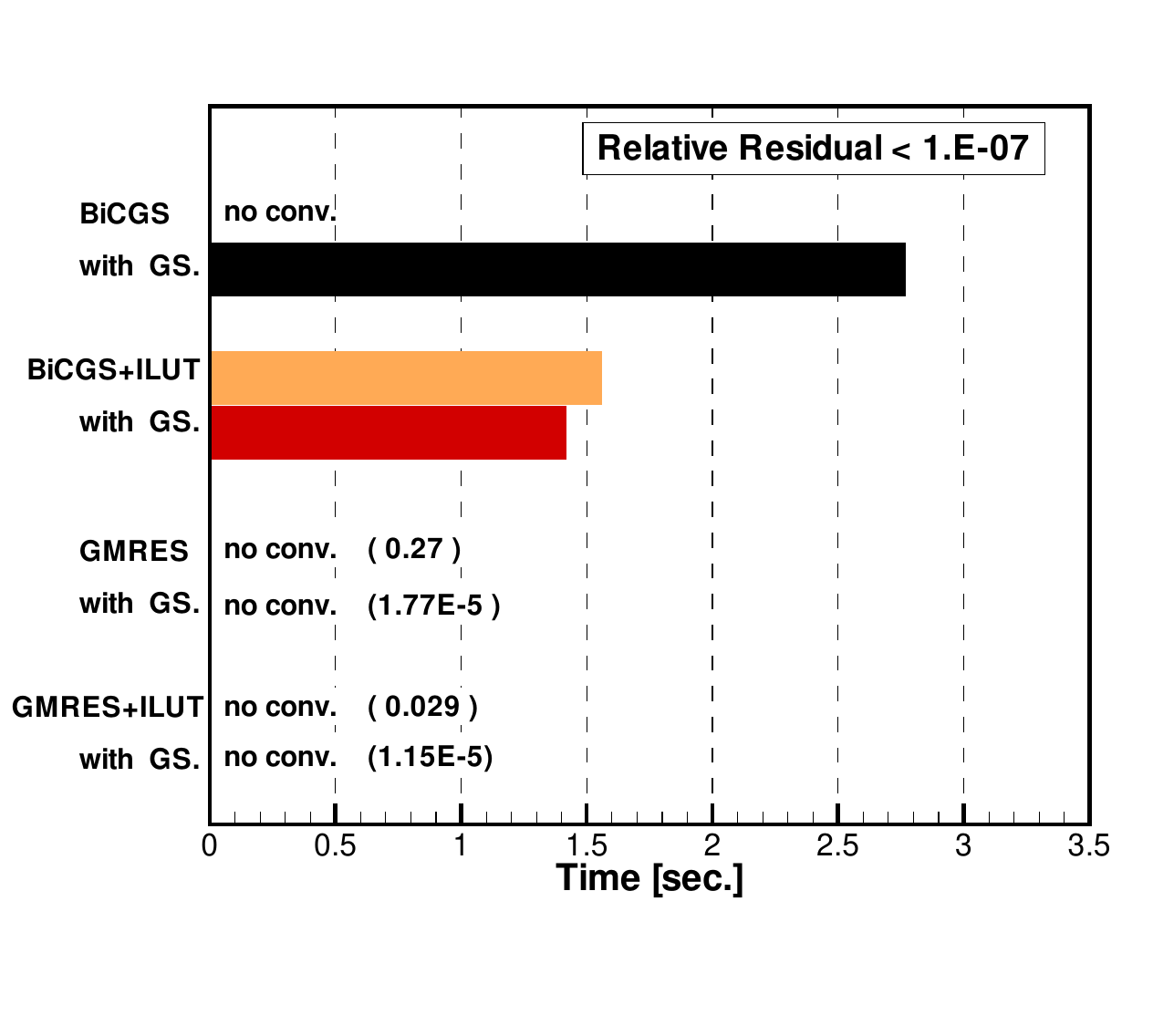}
&
\includegraphics[width=3.15in]{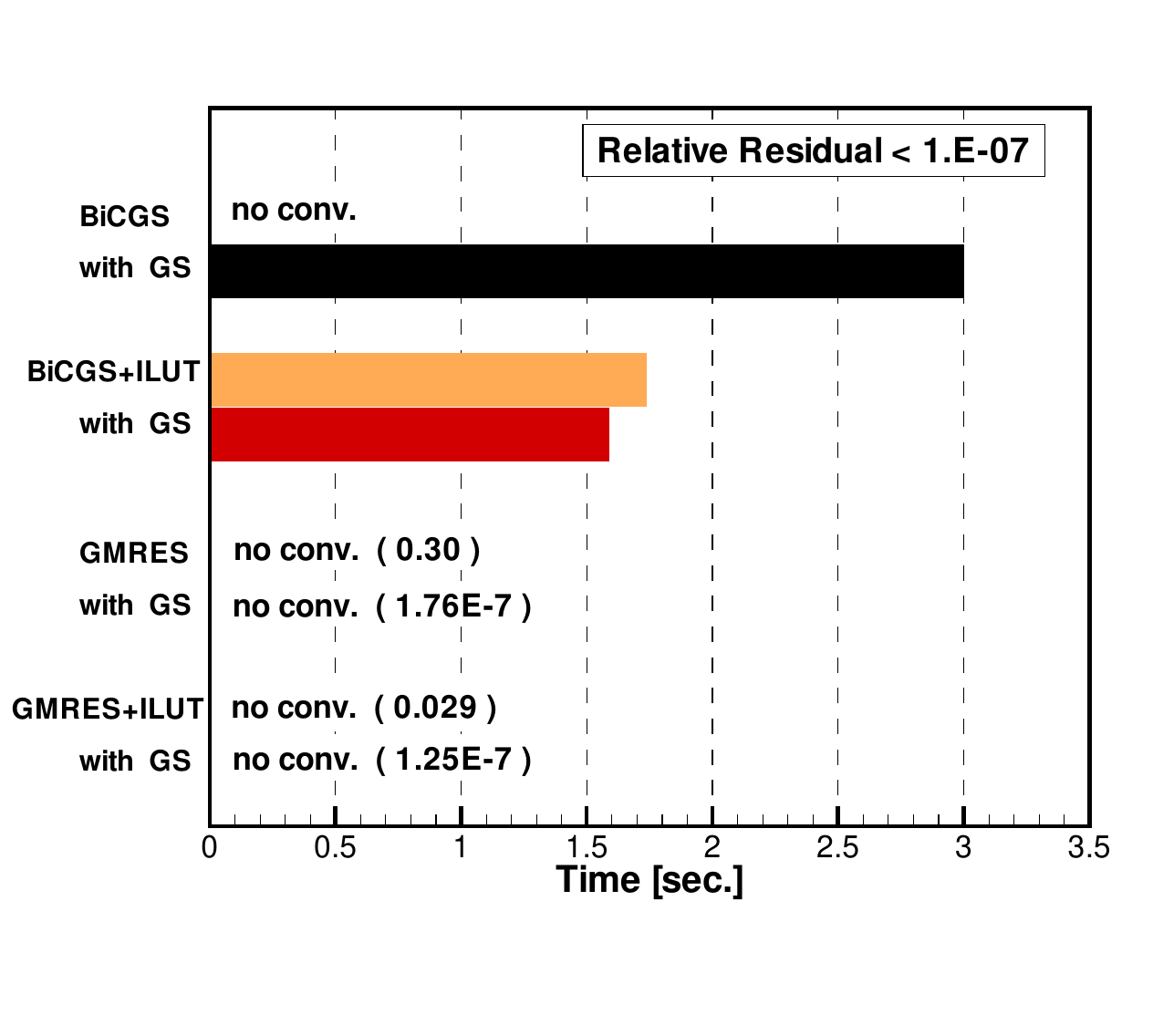}
\\[-.5in]
\includegraphics[width=3.15in]{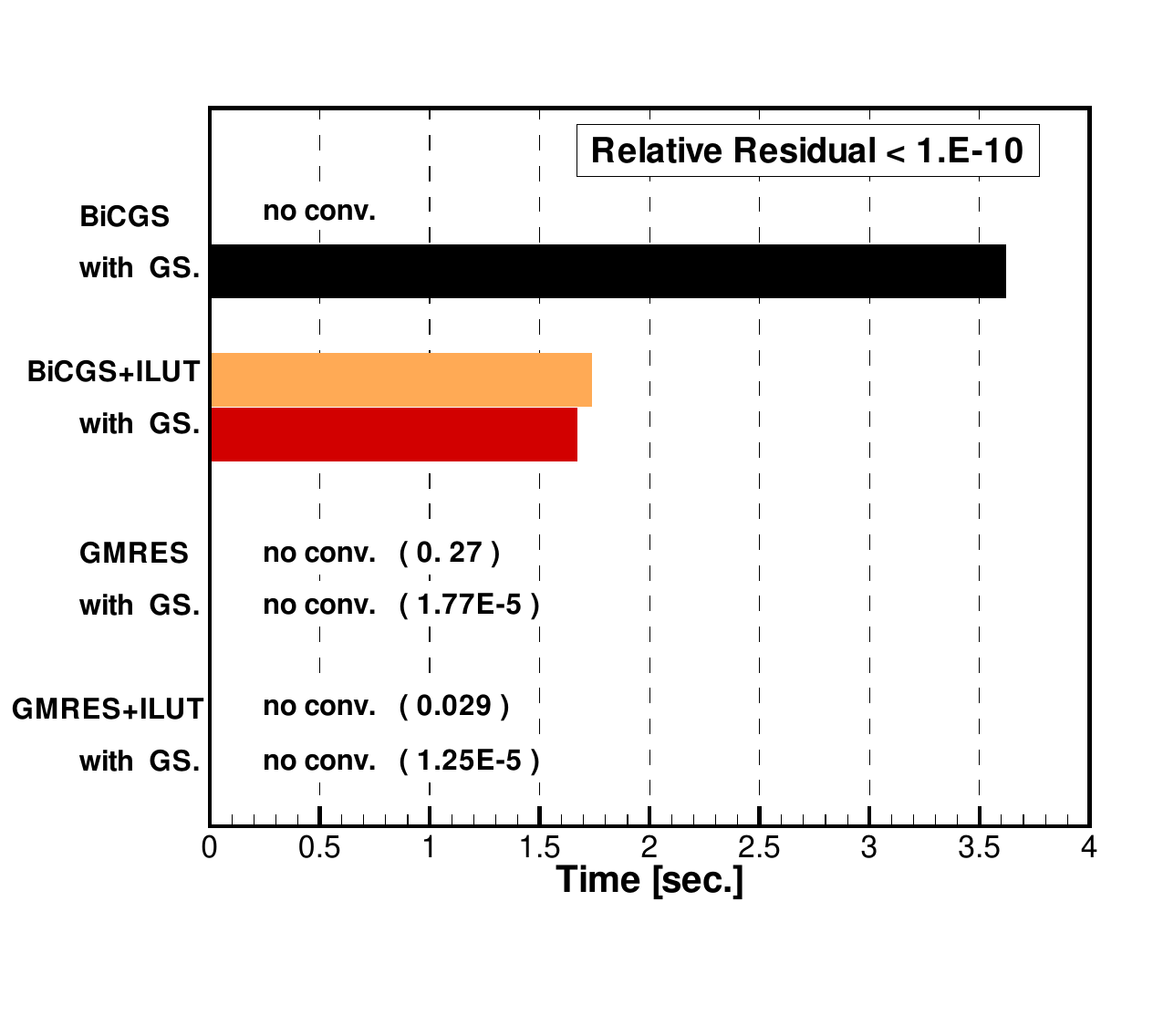}
&
\includegraphics[width=3.15in]{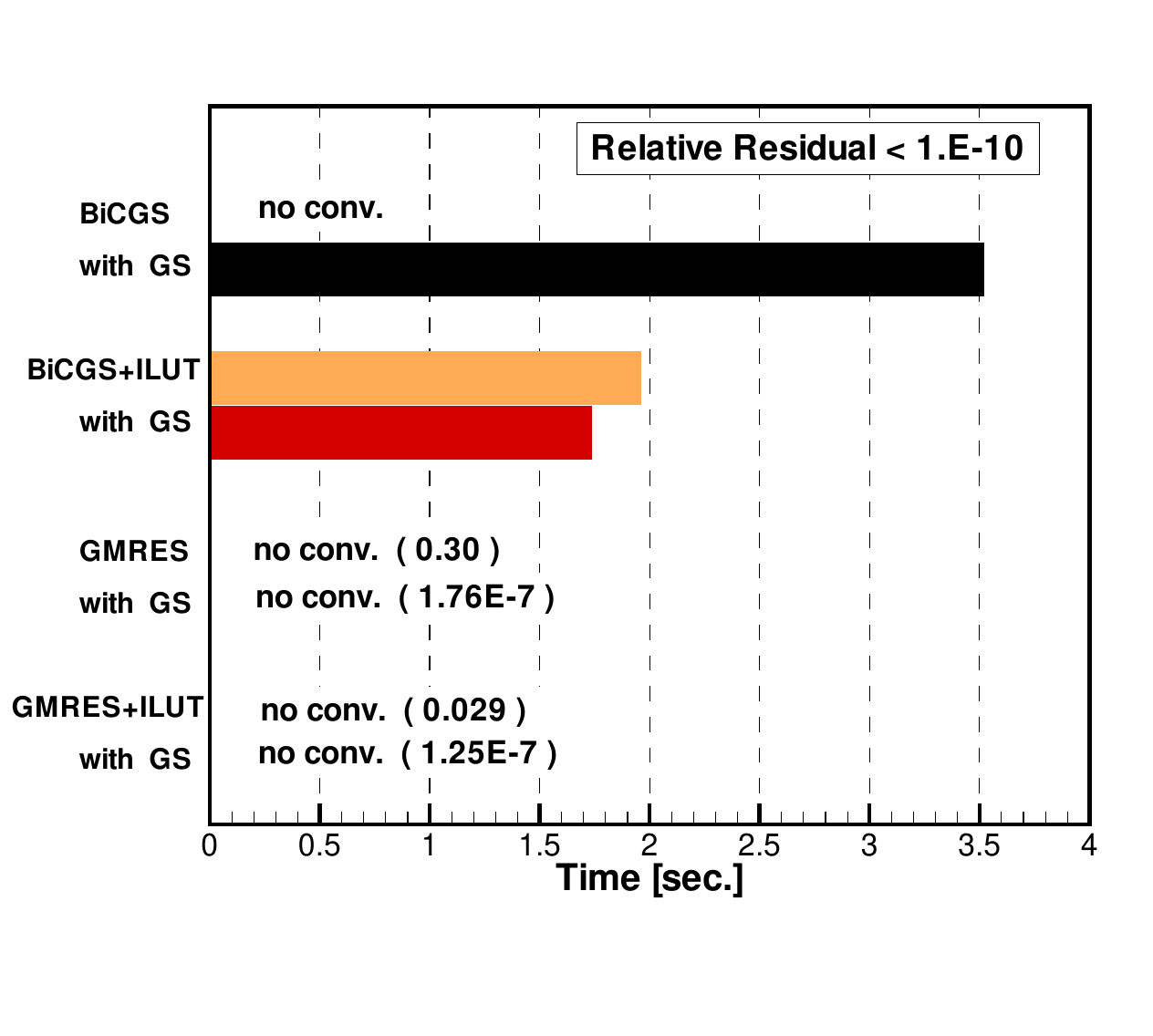} 
\end{tabular}
\vspace{-.5in}
\caption{Runtimes and convergence status for Problem 4, 
grid = $40\time40\time40$.}
\label{run4}
\end{figure}

\begin{figure}[!h]
\hspace{5pt}
\begin{tabular}{@{}c@{}c@{}}
Results for $D = 10^4$ & Results for $D = 10^6$
\\ \hline
\includegraphics[width=3.15in]{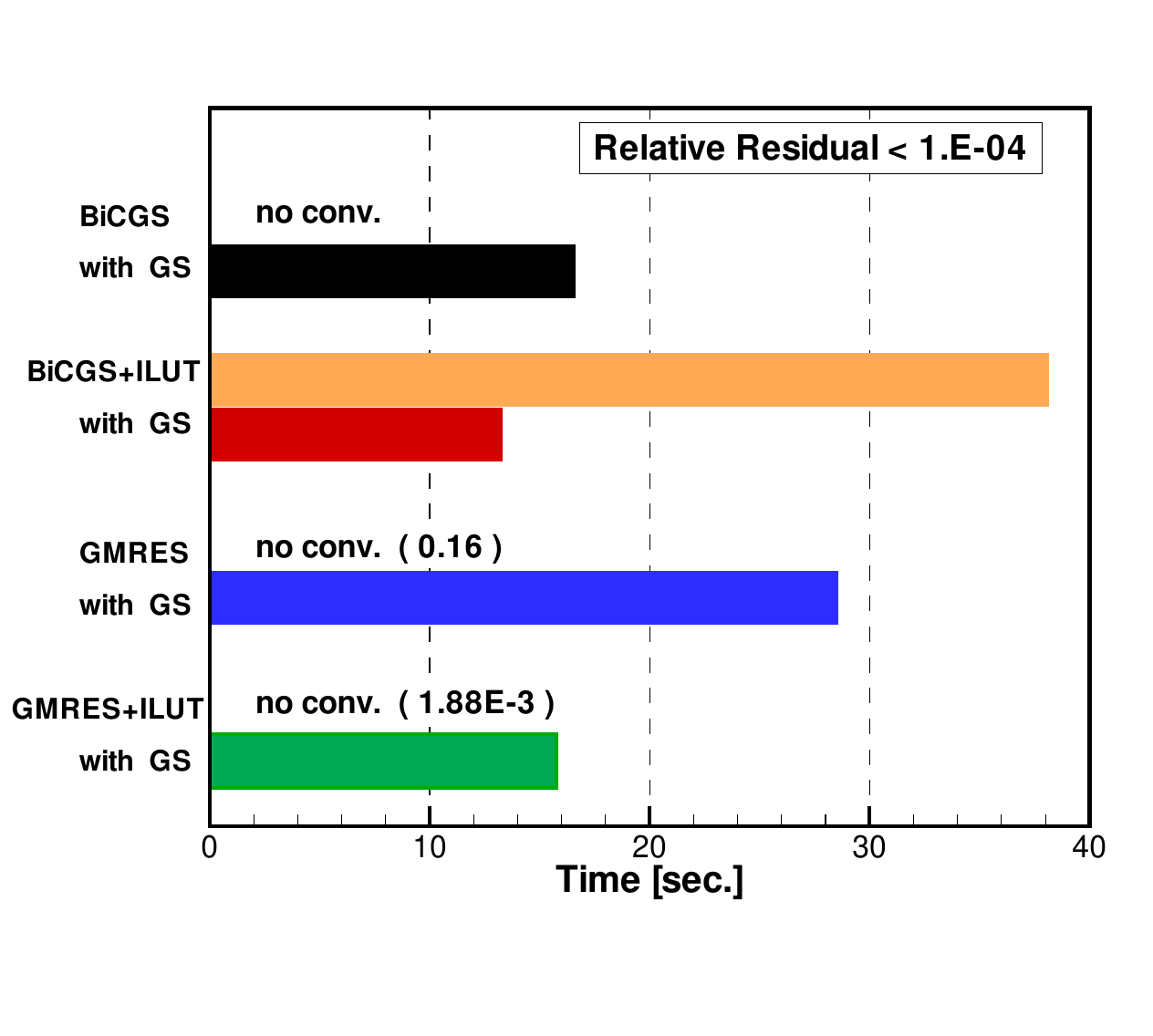}
&
\includegraphics[width=3.15in]{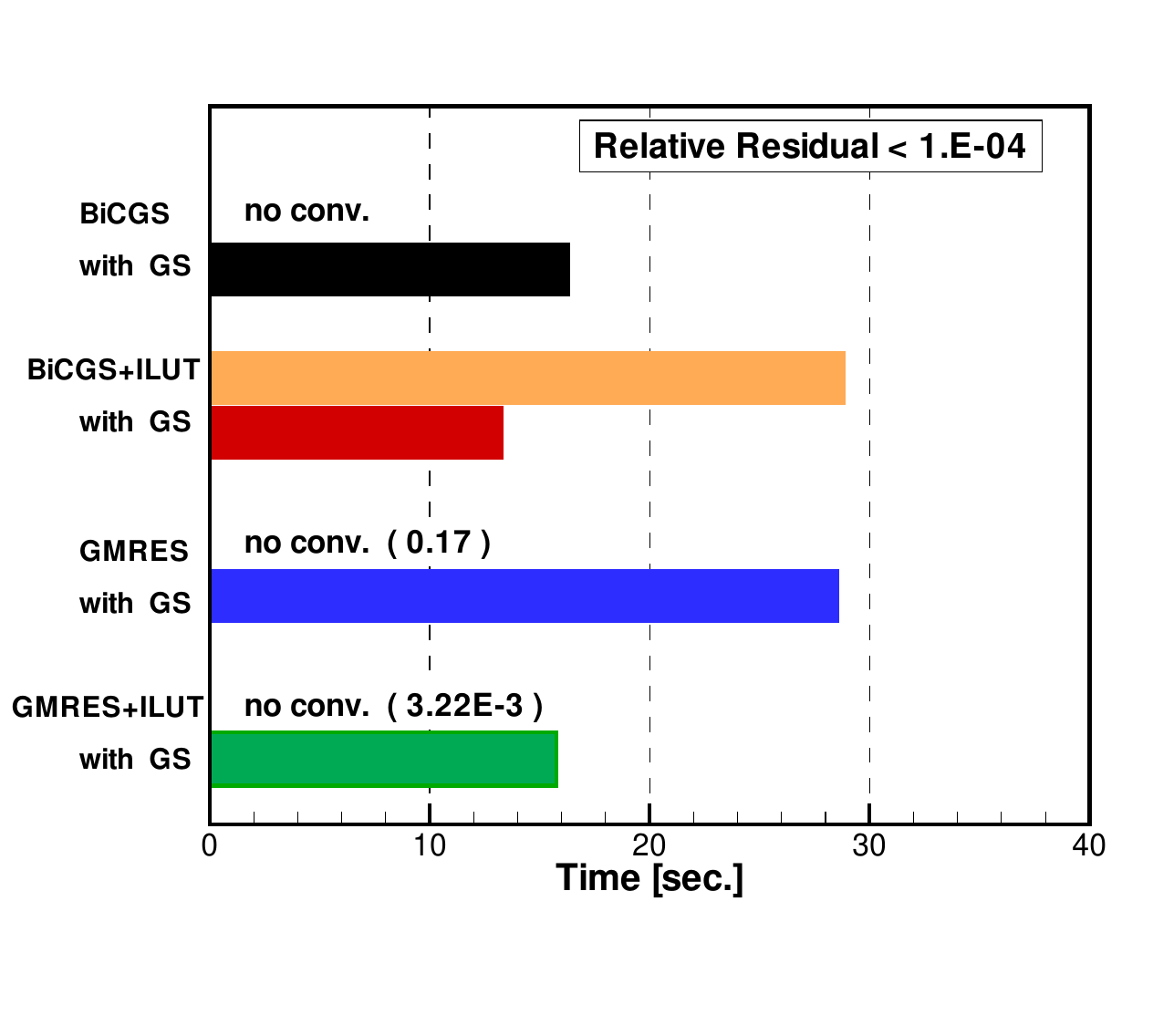}
\\[-.5in]
\includegraphics[width=3.15in]{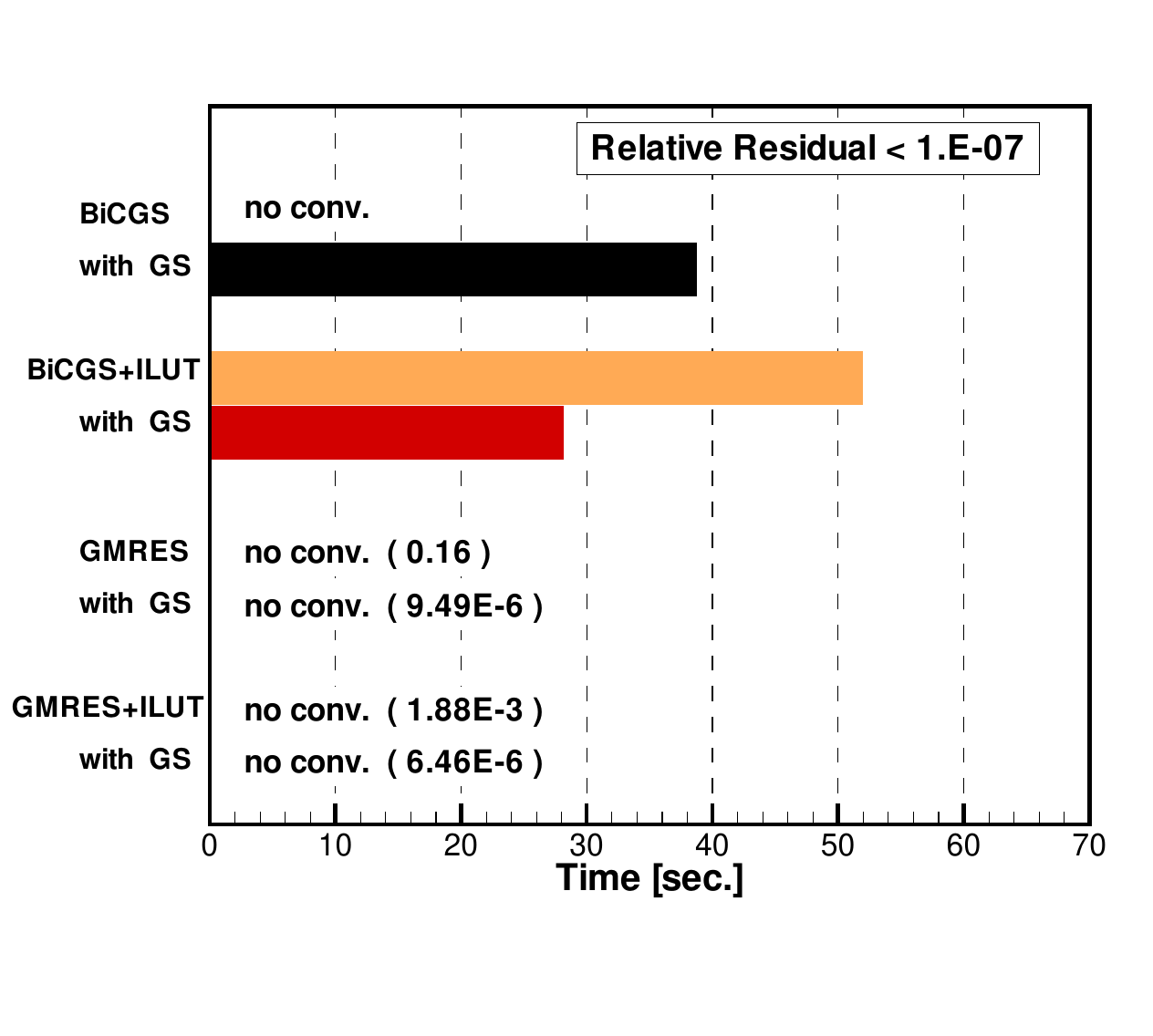}
&
\includegraphics[width=3.15in]{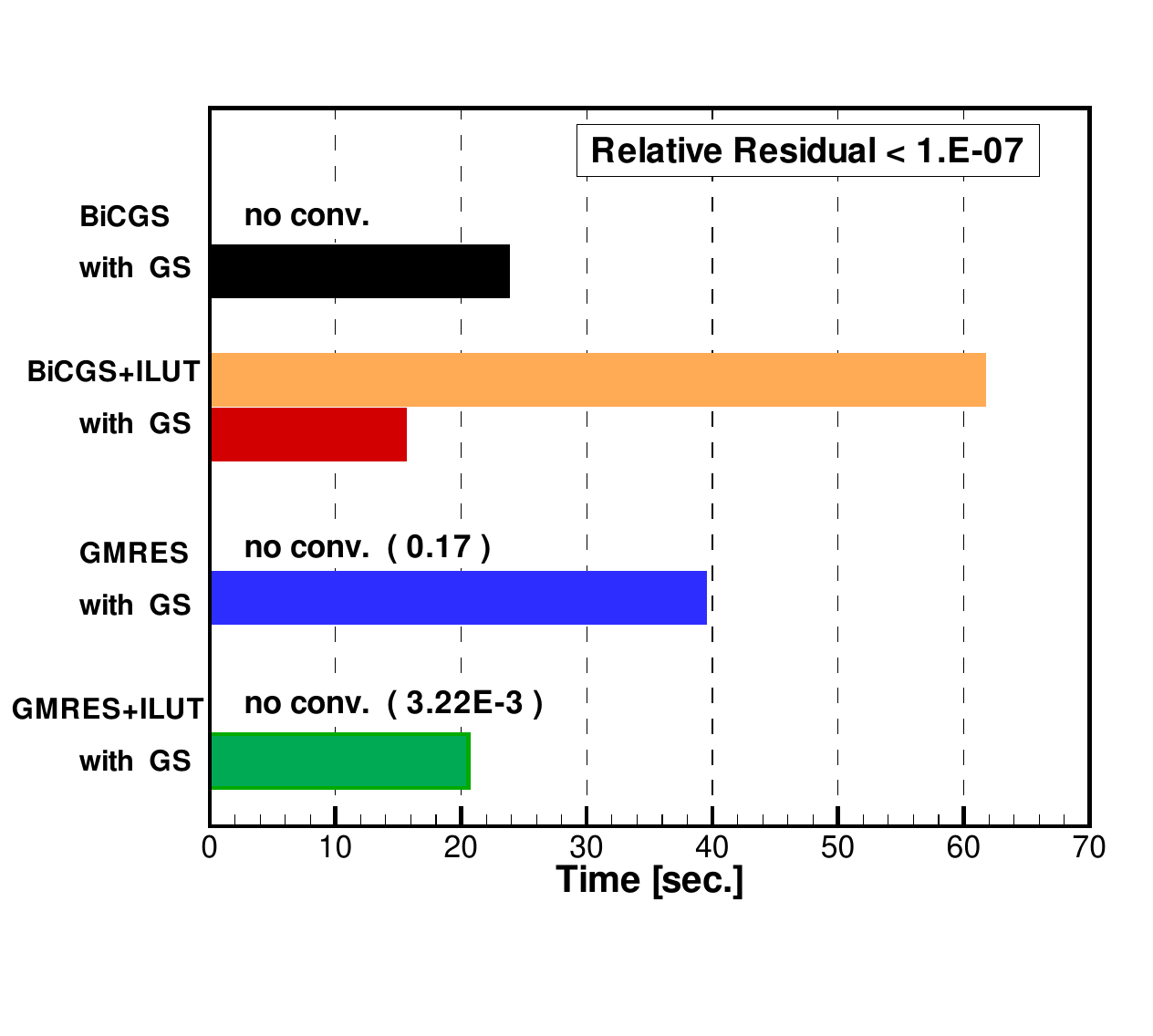}
\\[-.5in]
\includegraphics[width=3.15in]{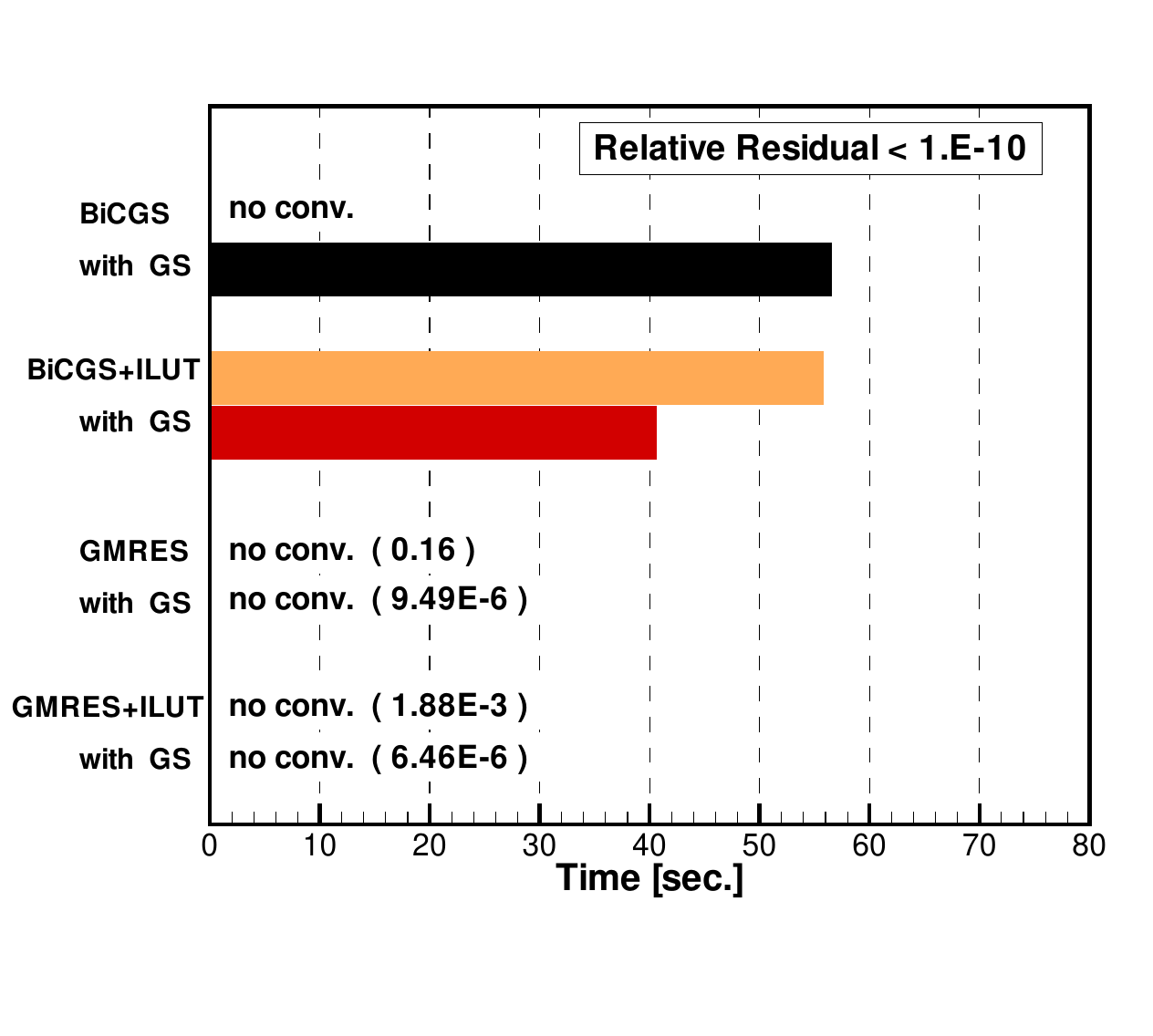}
&
\includegraphics[width=3.15in]{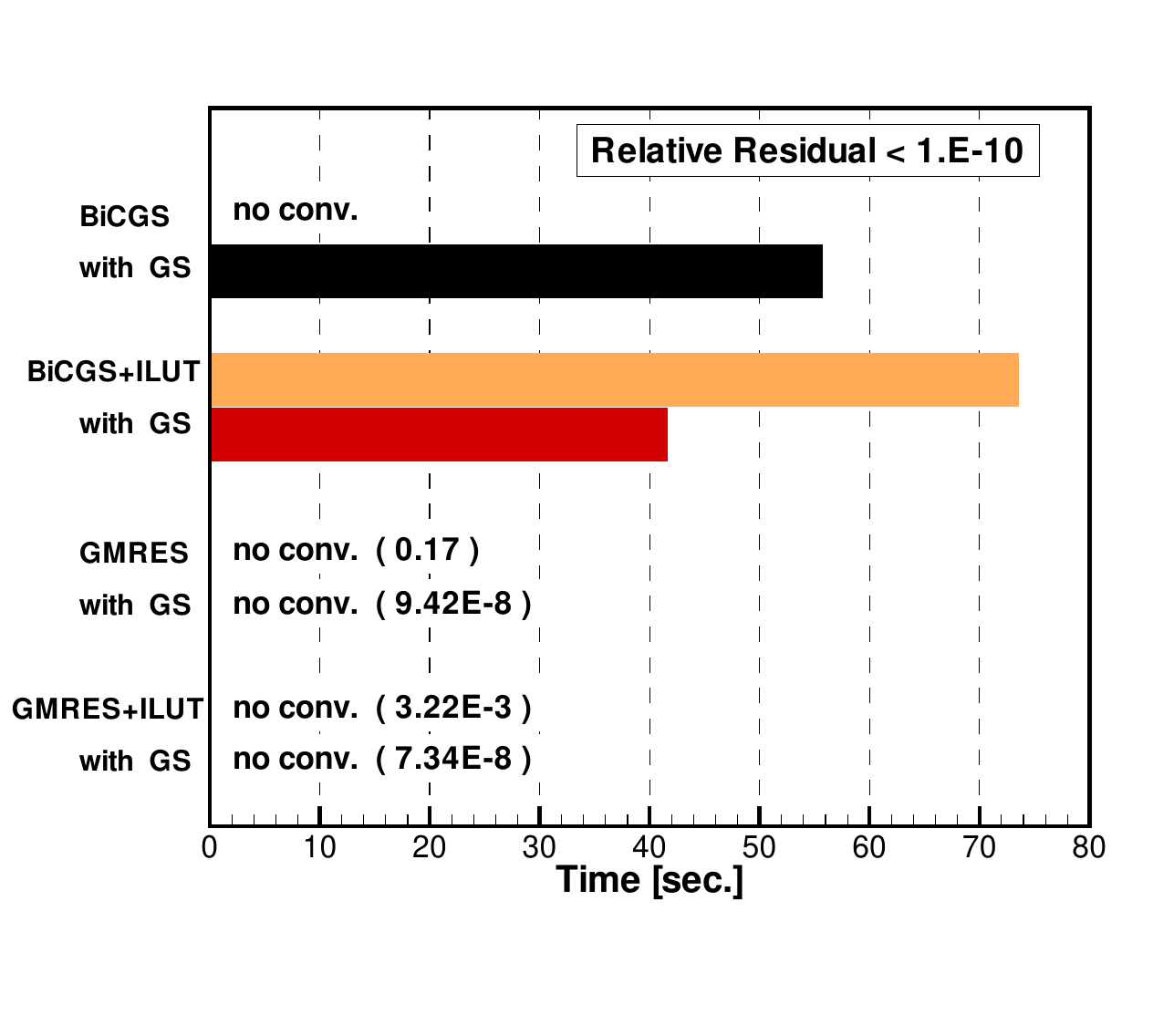}
\end{tabular}
\vspace{-.5in}
\caption{Runtimes and convergence status for Problem 4,
grid = $80\time80\time80$.}
\label{run4a}
\end{figure}

Figures \ref{run4} and \ref{run4a} show bar-plots of the runtimes 
for the two grid sizes, for $D=10^4$ and $D=10^6$, for the three 
levels of accuracy.  In cases of stagnation, the figures show (in 
parentheses) the relative residual achieved before stagnation.

The results can be summarized as follows.
\bi
\item Bi-CGSTAB (without ILUT) and GMRES (with and without ILUT)
did not converge in any of the cases.
\item GS enabled the convergence of Bi-CGSTAB in all cases,
and the convergence of GMRES (with and without ILUT) in the
low-accuracy cases and in one mid-accuracy case.
\item GS significantly improved the accuracy of GMRES (with and
without ILUT) by postponing the stage at which stagnation set in,
enabling convergence to much higher levels of accuracy.
\item For the $80\time80\time80$-grid, GS speeded up the convergence
of Bi-CGSTAB with ILUT quite significantly.  This was also true for
the coarser grid, but only in the low-accuracy case.
\item In the above cases, GS (by itself) was generally a better 
preconditioner than ILUT for Bi-CGSTAB.
\item In all cases with GS, the results for $D=10^6$ were either 
similar to or better than those for $D=10^4$.
\item In summary, GS was beneficial in all cases to all the 
algorithm/ILUT combinations, and in many cases, GS provided 
a very significant advantage.
\ei

Table \ref{tbl4a} provides the basic eigenvalue information for 
Problem 4 with $D=10^4$, for the original and the scaled matrices, 
and also for a continuous version with $D=10^4$ everywhere.  The 
grid size for this data was $12\time12\time12$.  The the last 
column shows the number of eigenvalues whose real part lies in the 
same interval as the origin (out of 100 intervals).  We can see 
that although GS doubled the condition number, it reduced the 
number of eigenvalues around the origin very significantly.

\begin{table}[!h]
\centering
\begin{tabular}{|l|c|c|c|c|}
\hline&&&&\\[-12pt]
{\bf Matrix} & $\lambda_{\min}$ & $\lambda_{\max}$ &
$\lambda_{\max} / \lambda_{\min}$ &
\parbox[c]{1.25in}{No.\ of eigenvalues around $x\!=\!0$\vspace{2pt}} \\
\hline&&&&\\[-12pt]
Original        & 1.45E-2 & 2.93E+3 & 2.02E+5 & 1131 \\
\hline&&&&\\[-12pt]
With GS      & 2.41E-6 & 1.00E+0 & 4.16E+5 & 25 \\
\hline&&&&\\[-12pt]
Cont.\ coef.\ ($D\!=\!10^4$) & 3.61E+0 & 6.18E+4 & 1.71E+4 & 42 \\
\hline
\end{tabular}
\caption{Basic eigenvalue information for Problem 4.}
\label{tbl4a}
\end{table}

\begin{figure}[!h]
\begin{tabular}{@{}cc@{}}
\hspace{.1in}
\includegraphics[width=3in]{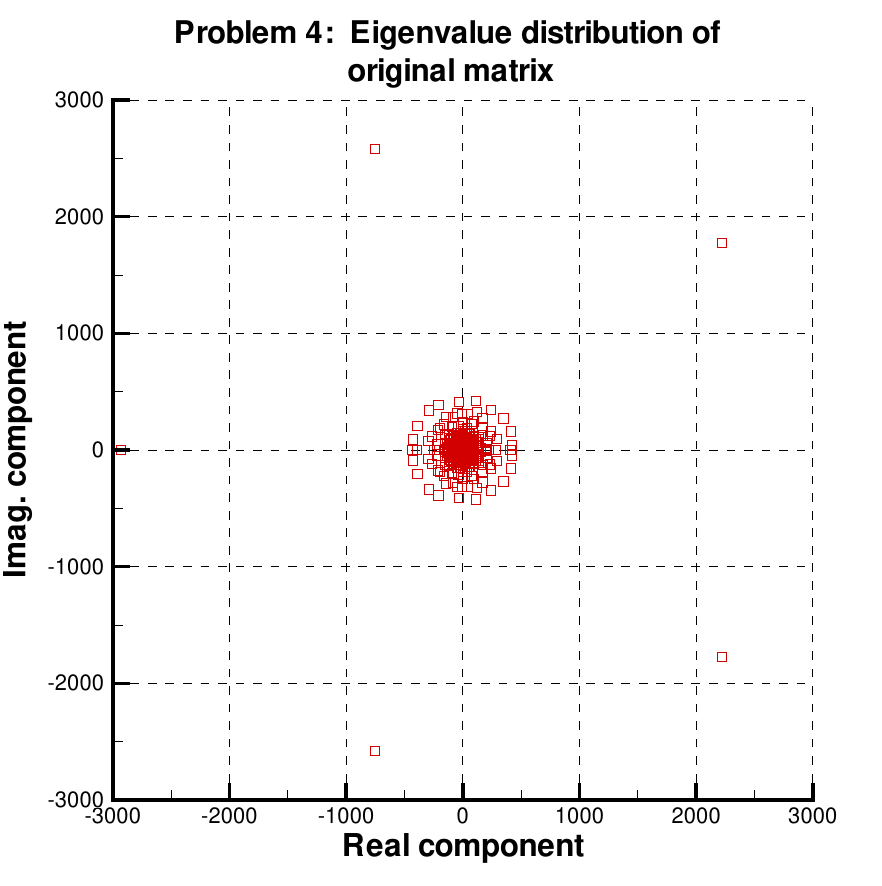}
&
\includegraphics[width=3in]{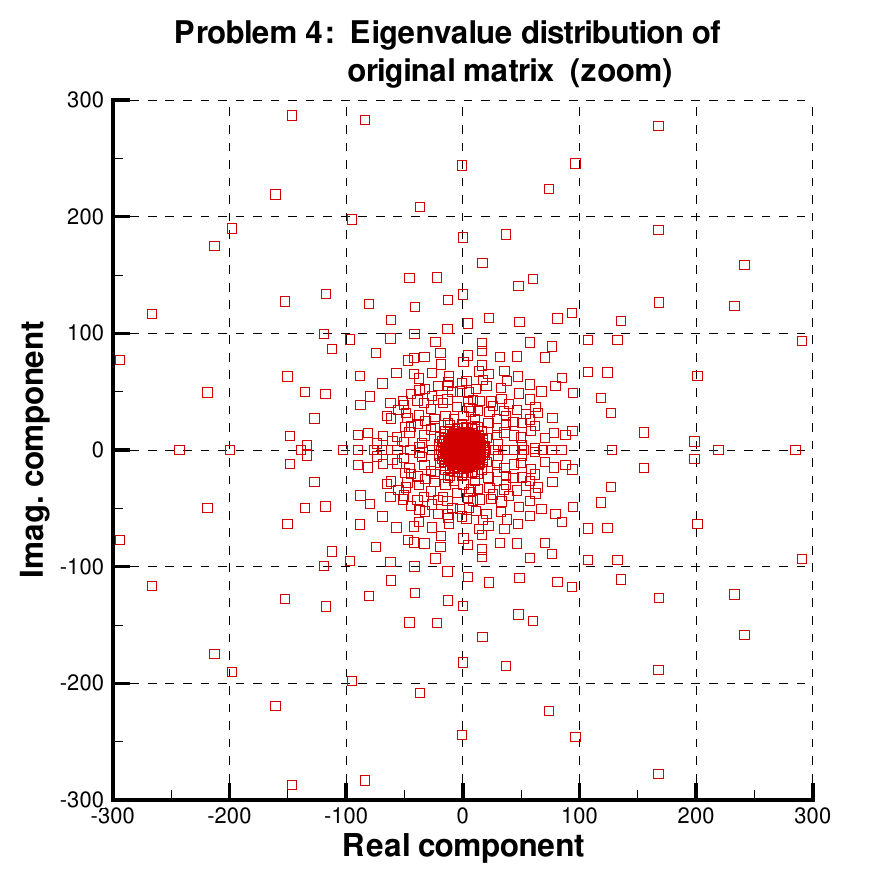} \\
\end{tabular}
\caption{Eigenvalue distribution for Problem 4, with a zoom to the region
$[-300,300]^2$.}
\label{eigen4}
\end{figure}

\begin{figure}[!h]
\begin{tabular}{@{}cc@{}}
\hspace{.1in}
\includegraphics[width=3in]{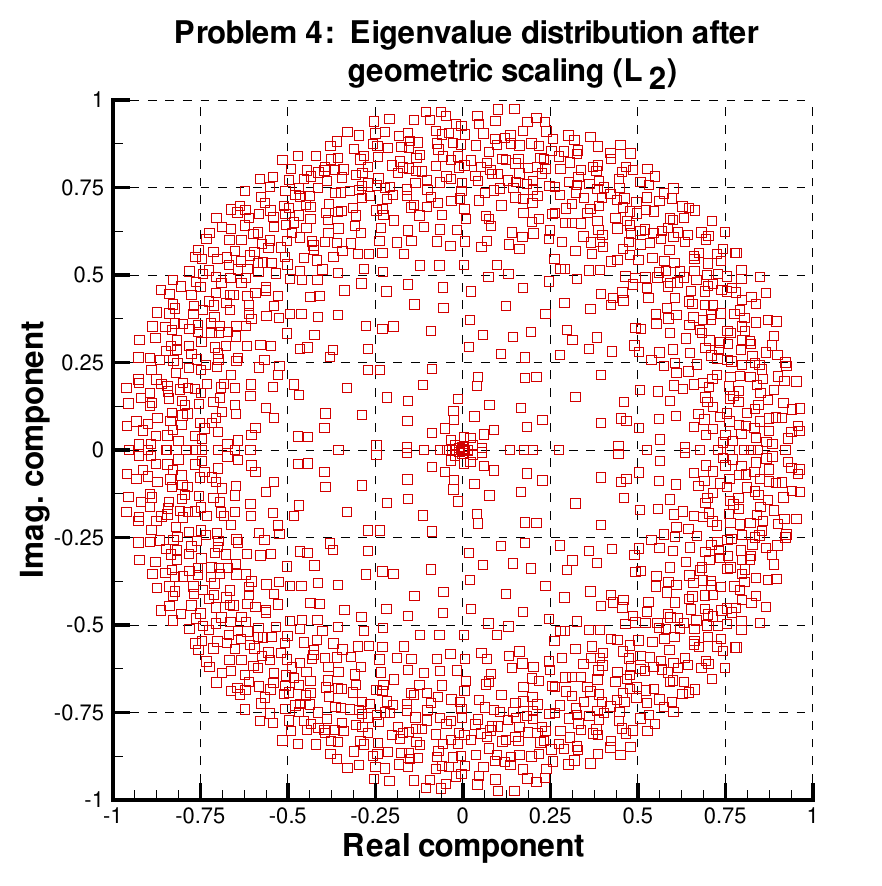}
&
\includegraphics[width=3in]{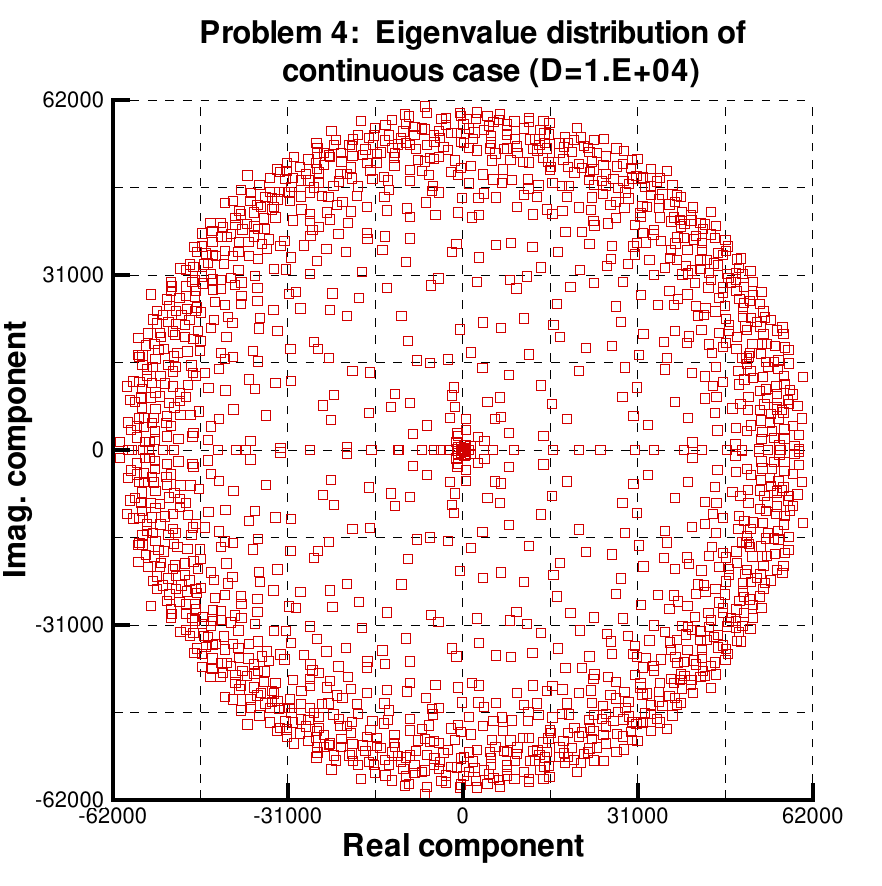} \\
\end{tabular}
\caption{Eigenvalue distribution for Problem 4, for the scaled and the 
continuous cases.}
\label{eigen4a}
\end{figure}

Figure \ref{eigen4} shows the distribution of the eigenvalues for the
original Problem 4, and a zoom to the region $[-300,300]^2$.  We can 
see that the eigenvalues are very concentrated around the origin.  
Figure \ref{eigen4a} shows the distribution of the eigenvalues for the 
scaled and the continuous ($D\!=\!10^4$) cases.  We can see that these
distributions look quite similar, with eigenvalues concentrated around 
the perimeter.

\begin{figure}[!h]
\centering
\includegraphics[width=4.5in]{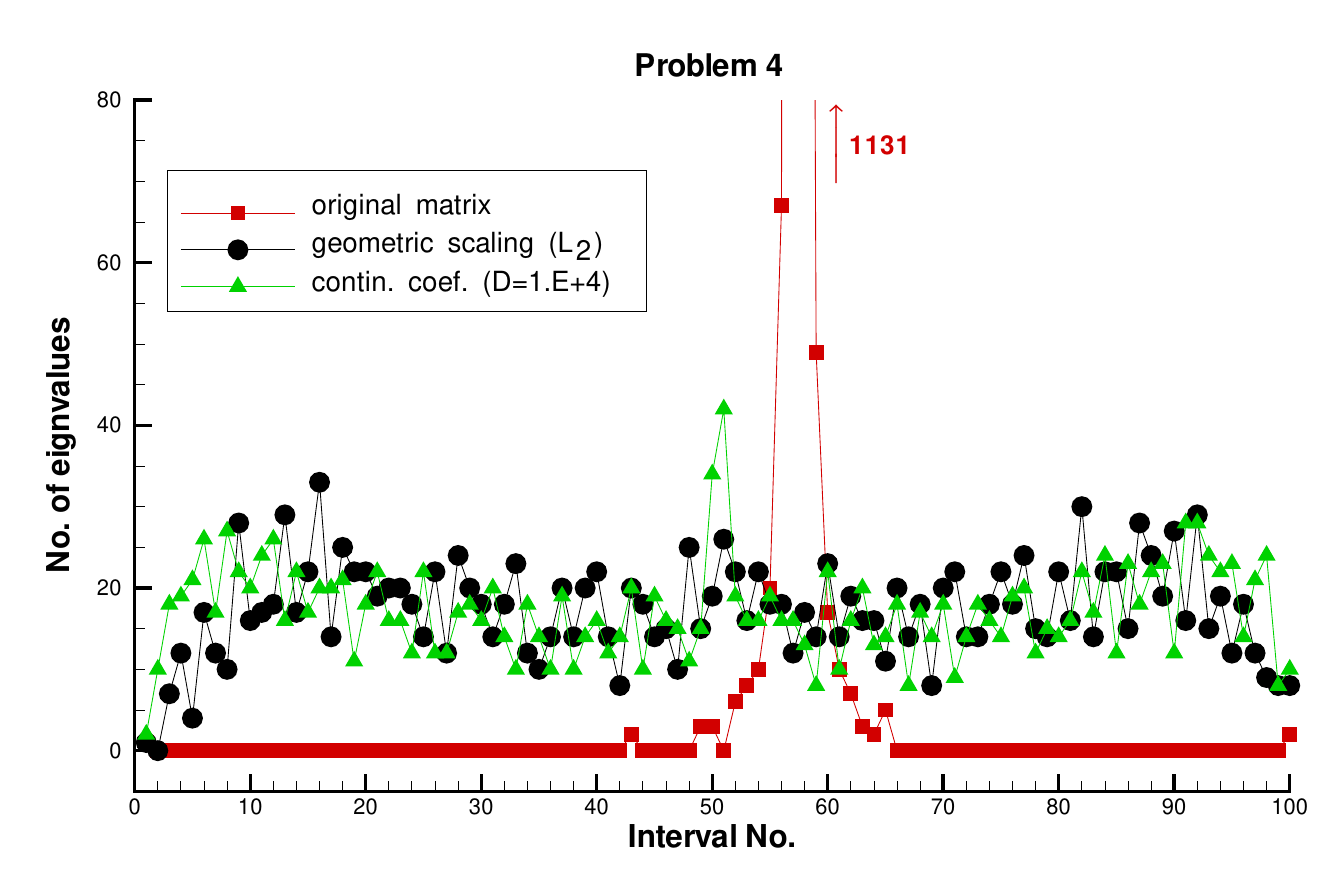}
\vspace{-.1in}
\caption{Histogram of the real part of the eigenvalues for Problem 4,
for the original, the scaled, and the continuous cases.}
\label{dist4}
\end{figure}

Figure \ref{dist4} provides a histogram of the real part of the eigenvalues
of Problem 4, for the original, the scaled, and the continuous ($D=10^4$) 
cases.  The histogram provides an additional view of how the eigenvalues 
are distributed in the three cases.

GS is proposed as being useful for discontinuous coefficients when 
the convection terms are small to moderate.  In order to demonstrate
this, we considered Problem 4 with $D=10^4$, a grid size of
$40\time 40\time 40$, and varying convection terms.  The previous 
results were based on convection terms of 100, so we also tested 
this case with convection terms of 200, 500, and 1000.  The results, 
which  are summarized in Table \ref{limit} below, show how the 
usefulness of GS degrades as the convection terms are increased.  

\begin{table}[!h]
\centering
\begin{tabular}{|l|c|c|c|c|}
\hline
&&&&\\[-12pt]
{\bf Method / Convection:} & {\bf 100} & {\bf 200} &
{\bf 500} & {\bf 1000} \\
\hline
&&&&\\[-12pt]
Bi-CGSTAB & --- & --- & --- & --- \\
with GS   & $10^{-10}$ & $10^{-10}$ & $10^{-10}$ & --- \\
\hline
&&&&\\[-12pt]
Bi-CGSTAB+ILUT & $10^{-10}$ & $10^{-4}$ & --- & --- \\
with GS        & $10^{-10}$ & $10^{-4}$ & --- & --- \\
\hline
&&&&\\[-12pt]
GMRES & --- & --- & --- & --- \\
with GS & $10^{-4}$ & $10^{-4}$ & --- & --- \\
\hline
&&&&\\[-12pt]
GMRES+ILUT & --- & --- & --- & --- \\
with GS    & $10^{-4}$ & --- & --- & --- \\

\hline
\multicolumn{5}{|l|}{}\\[-12pt]
\multicolumn{5}{|l|}{Note: `---' means no convergence. The numbers} \\
\multicolumn{5}{|l|}{indicate which relative error goal was obtained.} \\
\hline
\end{tabular}
\caption{Degradation of the various methods as the convection
terms are increased.}
\label{limit}
\end{table}

Note in particular the difference between Bi-CGSTAB without ILUT
and Bi-CGSTAB with ILUT: it turns out that without ILUT, GS enables
convergence to $10^{-10}$ even with convection = 500, but with ILUT,
degradation already appears with convection = 200; this degradation
is independent of the use of GS.  


\section{Conclusions and further research}
\label{conclusions}

In this paper we have reported on a very simple technique for 
improving the convergence properties of algorithms on certain 
linear systems with discontinuous coefficients.  The technique, 
called geometric scaling (GS), consists of simply dividing each 
equation by the $L_p$-norm of its vector of coefficients.  
GS, with $p=2$, was tested on four nonsymmetric problems derived 
from convection-diffusion elliptic PDEs, with small to moderate 
convection terms.  Bi-CGSTAB and restarted GMRES, with and 
without ILUT, were tested on the four problems with and without 
GS.  Normally, such problems are solved by domain decomposition 
(DD) techniques, but these can be difficult to implement if the 
boundaries between the subdomains have a complicated geometry 
or the grid is unstructured.  The problems were taken from (or 
based on) well-known examples from the literature, and included 
two- and three-dimensional cases, with Dirichlet and mixed 
Dirichlet/Neumann boundary conditions.  One problem had a 
complicated geometry.  Three different convergence goals were
prescribed: relative residual $\le 10^{-4}, 10^{-7}, 10^{-10}$.

Table \ref{tbl_sum} summarizes the convergence behavior of the four
different algorithm/preconditioner combinations, with and without 
GS, on the four different problems, for the three convergence goals.
The results indicate that GS was very useful in improving the 
convergence status of the different solution methods on the tested 
problems.  With one exception, when the solution method converged 
on a problem with discontinuous coefficients, GS speeded up the 
convergence.  Eigenvalue distribution maps show that when there 
is a strong concentration of values around the origin, GS reduces 
this concentration very significantly and ``pushes'' the values 
away from the origin.

\begin{table}[!h]
\centering
\begin{tabular}{|l|c|c|c|c|}
\hline
&&&&\\[-12pt]
{\bf Method} & {\bf Problem 1} & {\bf Problem 2} & 
{\bf Problem 3} & {\bf Problem 4} \\ 
\hline 
&&&&\\[-12pt]
Bi-CGSTAB & $-~-~-$ & $-~-~-$ & $-~-~-$ & $-~-~-$ \\ 
with GS   & $+~+~+$ & $7.2\time10^{-10}$ &  $-~-~-$ & $+~+~+$ \\
\hline 
&&&&\\[-12pt]
Bi-CGSTAB+ILUT & $+~+~+$ & $+~+~*$ & $+~+~+$ & $+~+~+$ \\
with GS        & $\!+~\,*~~*$ & $\!~*~*~+$ & $*~~*~~*$ & $*~~*~~*$ \\
\hline
&&&&\\[-12pt]
GMRES & $3.8\time 10^{-2}$ & 1.37 & $-~-~-$ & 0.27 \\
with GS & $1.1\time 10^{-5}$ & $2.2\time 10^{-5}$ & $-~-~-$ & 
$1.8\time 10^{-5}$ \\
\hline
&&&&\\[-12pt]
GMRES+ILUT & $3.9\time 10^{-3}$ & 0.90 & $-~-~-$ & 0.29 \\
with GS    & $1.1\time 10^{-5}$ & $1.9\time 10^{-5}$ & $+~+~+$ & 
$1.25\time 10^{-5}$ \\
\hline
\multicolumn{5}{|l|}{}\\[-12pt]
\multicolumn{5}{|l|}{Note: `$-$' means no convergence, `$+$' 
means convergence, `$*$' means} \\
\multicolumn{5}{|l|}{better convergence.  The numbers indicate the 
best relative error obtained.} \\
\hline
\end{tabular}
\caption{Summary of convergence of the different methods on the four 
problems, for the three convergence goals 
($10^{-4}, 10^{-7}$, and $10^{-10}$).}
\label{tbl_sum}
\end{table}

This study is only a first report on this subject, and much work 
remains to be done.  On the theoretical side, we have seen that 
GS improves the eigenvalue distribution, but further analysis is
still needed.  On the practical side, GS needs to be tested in 
conjunction with other algorithms and preconditioners, on various 
other difficult problems.

A natural question that arises is how does GS compare with DD
techniques?  Over the years, these methods have achieved a high 
level of efficiency, and a head-on runtime comparison is very 
much problem-dependent and a topic for further research.
However, an even more interesting topic is the combination of
the two methods: apply GS to the equations and then apply some 
standard DD method; hopefully, on some problems, GS could also 
improve the convergence properties of some DD approaches.

GS can also help the parallelization of solution methods in 
several ways.  Firstly, GS itself is a parallel computational 
step.  Secondly, it enables the partitioning of a domain into 
subdomains along boundaries that do not necessarily follow the 
physical boundaries of the problem; this way, better load 
balancing can be achieved.  Thirdly, as seen in some of the 
cases, GS enabled the convergence of algorithms that are 
inherently parallel without the need for a preconditioner 
(such as ILUT) that may not be ideal for parallelism.

Another topic for further research is the application of the
(sequential) CGMN algorithm \cite{Bjorck79,Gordon08} and its
block-parallel version CARP-CG \cite{Gordon08a} on problems 
that have discontinuous coefficients and are also strongly 
convection-dominated.  These algorithms have already been 
shown to be very effective on convection-dominated problems 
and initial experiments with discontinuous coefficients have 
yielded good results.  The reason for this may be that these 
two methods are essentially conjugate-gradient acceleration
of the Kaczmarz algorithm, and as such, GS (with the 
$L_2$-norm) is already inherent in them.  


\bibliography{}
\end{document}